\documentclass[reprint,aps,prr,twocolumn,superscriptaddress]{revtex4-1}
\usepackage{mhchem}
\usepackage[paperheight=11in,paperwidth=8.25in,margin=0.67in]{geometry}
\usepackage{amsmath}
\usepackage{amssymb}
\usepackage{graphicx,color}
\usepackage{url}
\usepackage{bbm}
\usepackage{algpseudocode}
\usepackage{algorithm}

\usepackage{amsthm}

\begin{document}
\title{
Quantum computation of molecular response properties
}
\author{Xiaoxia Cai}\affiliation{Key Laboratory of Theoretical and Computational Photochemistry, Ministry of Education, College of Chemistry, Beijing Normal University, Beijing 100875, China}
\author{Wei-Hai Fang}\affiliation{Key Laboratory of Theoretical and Computational Photochemistry, Ministry of Education, College of Chemistry, Beijing Normal University, Beijing 100875, China}
\author{Heng Fan}\affiliation{Institute of Physics, Chinese Academy of Sciences, Beijing 100190, China}
\author{Zhendong Li}\email{zhendongli@bnu.edu.cn}
\affiliation{Key Laboratory of Theoretical and Computational Photochemistry, Ministry of Education, College of Chemistry, Beijing Normal University, Beijing 100875, China}

\begin{abstract}
Accurately predicting response properties of molecules such as the dynamic polarizability and hyperpolarizability
using quantum mechanics has been a long-standing challenge with widespread applications in material and drug design. Classical simulation techniques in quantum chemistry are hampered by the exponential growth of the many-electron Hilbert space as the system size increases.
In this work, we propose an algorithm for computing linear and nonlinear molecular response properties
on quantum computers, by first reformulating the target property into a symmetric expression
more suitable for quantum computation via introducing a set of auxiliary quantum states, and then determining
these auxiliary states via solving the corresponding linear systems of equations on quantum computers.
On one hand, we prove that using the quantum linear system algorithm [Harrow et al., Phys. Rev. Lett.
103, 150502 (2009)] as a subroutine the proposed algorithm scales only polynomially
in the system size instead of the dimension of the exponentially large Hilbert space, and hence
achieves an exponential speedup over existing classical algorithms.
On the other hand, we introduce a variational hybrid quantum-classical variant of the
proposed algorithm, which is more practical for near-term quantum devices.
\end{abstract}
\maketitle

{\it Introduction.} How molecules response upon the action of external fields
determines the properties of materials. For weak external fields,
the response is fully characterized by the linear and nonlinear response functions\cite{boyd2003nonlinear,norman2018principles},
such as the polarizability tensor $\alpha_{ij}(\omega)$ and hyperpolarizability
$\beta_{ijk}(\omega_1,\omega_2)$ ($i,j,k\in\{x,y,z\}$).
The dynamic polarizability $\alpha_{ij}(\omega)$ describes how the
dipole moment of a molecule responses to an oscillating electric field
to the leading order, and can be linked to the photoabsorption cross section,
while the first-order hyperpolarizability describes nonlinear response processes such as
second-harmonic generation in nonlinear optical materials.
Besides, these response functions are also the key to understand
intermolecular interactions. Notably, the
van der Waals $C_6$ coefficients,
which are of paramount importance in quantifying the dispersion interaction between
drug molecules and proteins in drug design,
can be computed from dynamic polarizabilities
at imaginary frequencies via the Casimir-Polder integral\cite{casimir1948influence}.

Developing reliable quantum mechanical methods for accurately predicting molecular response properties
has been one of the major challenges in quantum chemistry\cite{norman2018principles,helgaker2012recent,norman2018simulating}.
The full configuration interaction (FCI)\cite{szabo2012modern,olsen1985linear,koch1991analytical}, also known as
the exact diagonalization method, represents the most accurate method within a molecular orbital basis set,
however, is limited to small molecular systems due to the exponential growth of the many-electron Hilbert
space as the system size increases.
Over the past decades, a plethora of approximate methods along with efficient algorithms
have been developed\cite{helgaker2012recent,norman2018simulating}.
Unfortunately, approximations adopted in these methods in order to describe the correlation
among electrons efficiently, such as the mean-field approximation\cite{szabo2012modern} or
approximate exchange-correlation functionals in
density functional theory\cite{cohen2008insights}, can sometimes fail miserably.
In particular, the strong electron correlation\cite{kent2018toward}, which is the root
for many fascinating phenomena in materials such as high-temperature
superconductivity, cannot be accurately accounted for by these approximate methods.
A satisfactory classical simulation method for predicting molecular response properties, which works
in all regime of electron correlation, is lacking.

Initially advocated by Feynman\cite{feynman1982simulating},
quantum computation shows a great promise for solving interacting fermion problems
in physics and chemistry\cite{lloyd1996universal,abrams1997simulation,abrams1999quantum,
ortiz2001quantum,somma2002simulating,georgescu2014quantum,cao2019quantum,yuan2020}.
The quantum phase estimation (QPE) algorithm\cite{kitaev1995quantum}
was applied to obtain the ground state energies of molecules with an exponential speedup
over the classical FCI\cite{aspuru2005simulated}.
It also allows to compute molecular static properties via energy derivatives\cite{kassal2009quantum,o2019calculating}.
While QPE has only been realized for two-electron systems\cite{lanyon2010towards,du2010nmr,wang2015quantum,o2016scalable}
due to the requirement of long circuit depth,
the variational quantum eigensolver (VQE)\cite{peruzzo2014variational,mcclean2016theory}
is more suitable for the noisy intermediate-scale quantum (NISQ)\cite{preskill2018quantum} devices.
Unlike QPE, its advantage over classical simulation techniques in quantum chemistry is still an open question
and being actively explored. Nevertheless, VQE has been experimentally demonstrated
on various platforms for small molecules
such as \ce{H2}, \ce{HeH+}, \ce{LiH} and \ce{BeH2}\cite{peruzzo2014variational,o2016scalable,shen2017quantum,kandala2017hardware,hempel2018quantum}.
In view of such progresses on the ground state problem, it is
a natural question to ask whether computing molecular response
properties, as the next logical step after computing the ground state,
will also benefit from quantum computation.

In this work, we propose an algorithm for computing molecular response properties
on quantum computers. While dynamical properties can alternatively be obtained
by Fourier transform of the corresponding correlation functions
in the time domain\cite{somma2002simulating,chiesa2019quantum,francis2020quantum}
determined from real-time Hamiltonian simulations,
analogous to the classical computation side\cite{norman2018principles,helgaker2012recent,norman2018simulating} it is highly
desirable to have a quantum algorithm
for computing a target response property such as $\alpha_{ij}(\omega)$
or $\beta_{ijk}(\omega_1,\omega_2)$ at given frequencies directly.
Because in many molecular applications\cite{norman2018principles,helgaker2012recent},
only a small range of frequencies is of interest,
including the simulations of (hyper)polarizabilities
at specific frequencies of applied electromagnetic fields\cite{koch1991analytical,norman2005nonlinear},
absorption spectra in an interested visible/ultraviolet/X-ray region\cite{norman2018simulating},
and multi-dimensional spectroscopies for studying couplings
between selected modes\cite{mukamel2000multidimensional}.
By reformulating the target property into a symmetric expression
with the help of a set of auxiliary quantum states, we convert
the most demanding part of computations
into linear systems of equations for determining these states, which can be solved on quantum computers
using quantum algorithms for linear systems of equations\cite{harrow2009quantum,ambainis2010variable,
clader2013preconditioned,childs2017quantum,subacsi2019quantum}
or variational hybrid quantum-classical algorithms\cite{xu2019variational,bravo2019variational}.
Depending on the subroutine employed for determining auxiliary states,
the resulting variant of the proposed algorithm can be considered as the analog of QPE or
VQE for molecular response properties. While the later variational hybrid
quantum-classical variant is more practical for near-term quantum devices,
we prove that in combination with the quantum linear system algorithm
invented by Harrow, Hassidim, and Lloyd
(HHL)\cite{harrow2009quantum}, the runtime complexity of the quantum variant of our algorithm
scales polynomially in the molecular system size, instead of the dimension of the exponentially large many-electron
Hilbert space. Thus, an exponential speedup can be achieved  
compared with the classical FCI-based approach\cite{olsen1985linear,koch1991analytical}, 
which laid down a firm foundation for the future application of quantum computation
in predicting molecular response properties.

{\it Theory.} For concreteness, we consider the calculation
of the polarizability $\alpha_{zz}(\omega)$ for a molecule under
a monochromatic electric field with optical frequency $\omega$
in the $z$-direction. The static polarizability will be obtained as
a special case where $\omega=0$.
Extensions to off-diagonal components of the polarizability tensor
as well as nonlinear response properties
are straightforward and will be discussed later.

Suppose initially without external fields, a molecule with $N$ electrons
is in the ground state $|\Psi_0\rangle$ of the second quantized
electronic Hamiltonian $\hat{H}_0$, expressed in an orthonormal molecular spin-orbital basis $\{\psi_p\}_{p=1}^{K}$
($K$ is proportional to the system size $N$) as
\begin{eqnarray}
\hat{H}_0 &=& \sum_{p,q=1}^{K}h_{pq}a_p^\dagger a_q + \frac{1}{2}
\sum_{p,q,r,s=1}^{K}h_{pqrs}a_p^\dagger a_q^\dagger a_s a_r, \label{eq:H0}
\end{eqnarray}
where $a^{(\dagger)}_p$ represents the fermionic annihilation (creation) operator, and $h_{pq}$ ($h_{pqrs}$) represent
the one-electron (two-electron) integrals.
The dynamic electric field in the dipole approximation is associated with the perturbation operator
\begin{eqnarray}
\hat{z} = \sum_{p,q=1}^{K}z_{pq}a_p^\dagger a_q,\label{eq:Vz}
\end{eqnarray}
where $z_{pq}\triangleq\langle\psi_p|z|\psi_q\rangle$ represent the
dipole-moment integrals. By the time-dependent perturbation theory,
the frequency-dependent polarizability $\alpha_{zz}(\omega)$
can be expressed in a sum-over-state (SOS) form\cite{boyd2003nonlinear,norman2018principles}
\begin{eqnarray}
\alpha_{zz}(\omega)
&=&
\sum_{n>0}\left[
\frac{\langle \Psi_0|\hat{z}|\Psi_n\rangle\langle\Psi_n|\hat{z}|\Psi_0\rangle}
{\omega_{n0}-(\omega+i\gamma)}\right.\nonumber\\
&&\left.+\frac{\langle \Psi_0|\hat{z}|\Psi_n\rangle\langle \Psi_n|\hat{z}|\Psi_0\rangle}
{\omega_{n0}+(\omega+i\gamma)}\right],\label{eq:SOS}
\end{eqnarray}
with $\gamma>0$ being a phenomenological damping parameter, which physically is associated with
the inverse lifetime of excited states.
Computing $\alpha_{zz}(\omega)$
allows to access important information of molecules such as the
transition dipole moments $\langle\Psi_0|\hat{z}|\Psi_n\rangle$ between
the ground state $|\Psi_0\rangle$ and the $n$-th excited state $|\Psi_n\rangle$,
as well as the associated excitation energy $\omega_{n0}\triangleq E_n-E_0$.
Moreover, the imaginary part of $\alpha_{zz}(\omega)$ is related with
the photoabsorption cross section $\sigma(\omega)\propto\omega\Im\alpha(\omega)$ for visible/ultraviolet/X-ray absorption spectra, which is one of the central quantities considered in designing functional materials.

In the standard FCI-based approach\cite{olsen1985linear,koch1991analytical}
for computing $\alpha_{zz}(\omega)$, Eq. \eqref{eq:SOS} is reformulated as
\begin{eqnarray}
\alpha_{zz}(\omega)&=&
\langle \Psi_0|\hat{z}|\Psi(\omega)\rangle+
\langle \Psi_0|\hat{z}|\Psi(-\omega)\rangle,\label{eq:alpha}
\end{eqnarray}
where the frequency-dependent response wavefunctions $|\Psi(\pm\omega)\rangle$
are obtained by solving the response equations
\begin{eqnarray}
\hat{Q}[\hat{H}_0-E_0\mp(\omega+i\gamma)]\hat{Q}|\Psi(\pm\omega)\rangle = \hat{Q}\hat{z}|\Psi_0\rangle,\label{eq:rspw}
\end{eqnarray}
with the projector $\hat{Q}=1-|\Psi_0\rangle\langle\Psi_0|$, in the full $N$-electron Hilbert space, and hence avoids
the need for determining all excited states in the SOS form \eqref{eq:SOS}.
The computational complexity of solving Eq. \eqref{eq:rspw} using
the best classical iterative algorithm\cite{saad2003iterative} for linear systems of equations
scales linearly in the dimension of the $N$-electron Hilbert space $D$.
For the molecular problem with $\hat{H}_0$ \eqref{eq:H0},
$D$ is exponential in $N$, e.g., $D=\binom{K}{N}$ with $K=2N$ for the half-filling case.
Therefore, like solving the ground-state eigenvalue problem, viz.,
$\hat{H}_0|\Psi_0\rangle=E_0|\Psi_0\rangle$,
this FCI-based approach scales exponentially in the system size $N$,
and in practical is limited to very small molecules (ca. $N\lesssim16$ assuming $K=2N$)
in routine quantum chemistry applications\cite{helgaker2012recent,norman2018simulating,koch1991analytical}.

Just as QPE and VQE have been applied to the ground state problem,
we attempt to utilize the advantage of quantum algorithms
for linear systems of equations\cite{harrow2009quantum,ambainis2010variable,
clader2013preconditioned,childs2017quantum,subacsi2019quantum,xu2019variational,bravo2019variational}
in computing molecular response properties.
However, while QPE can be applied readily to the ground state problem,
both Eqs. \eqref{eq:alpha} and \eqref{eq:rspw}
are not in a form that is amenable to compute on quantum computers directly,
due to the asymmetric form of each term in Eq. \eqref{eq:alpha} and the involvement
of the projector $\hat{Q}$. To resolve these two problems,
we introduce the notation
\begin{eqnarray}
\hat{A}(\pm\omega)\triangleq\hat{H}_0-E_0\mp(\omega+i\gamma)\label{eq:Aw}
\end{eqnarray}
for brevity and rewrite the first part of $\alpha_{zz}(\omega)$ \eqref{eq:alpha} as
\begin{eqnarray}
\langle \Psi_0|\hat{z}|\Psi(\omega)\rangle&=&\langle \Psi_0|\hat{z} \hat{Q}[\hat{Q}\hat{A}(\omega)\hat{Q}]^{-1}\hat{Q} \hat{z}|\Psi_0\rangle\nonumber\\
&=&
\langle \Psi_0|\hat{z} \hat{Q}\hat{A}^{-1}(\omega) \hat{z}|\Psi_0\rangle\nonumber\\
&=&
\langle \Psi_0|\hat{z}[\hat{A}^{\dagger}(\omega)]^{-1}
\hat{A}^\dagger(\omega)\hat{Q}
\hat{A}^{-1}(\omega)\hat{z}|\Psi_0\rangle,\label{eq:alpha1}
\end{eqnarray}
where the second equality follows from the spectral decompositions
$\hat{A}(\omega)=\sum_{n\ge0}|\Psi_n\rangle[\omega_{n0}-(\omega+i\gamma)]\langle\Psi_n|$
and $\hat{Q}=\sum_{n>0}|\Psi_n\rangle\langle\Psi_n|$, which immediately imply that
$\hat{A}(\omega)$ is invertible for $\gamma>0$ regardless of $\omega$, and
$\hat{Q}[\hat{Q}\hat{A}(\omega)\hat{Q}]^{-1}\hat{Q}=
\sum_{n>0}|\Psi_n\rangle[\omega_{n0}-(\omega+i\gamma)]^{-1}\langle\Psi_n|=
\hat{Q}\hat{A}^{-1}(\omega)\hat{Q}=\hat{Q}\hat{A}^{-1}(\omega)$.
To reach a symmetric expression, the identity $[\hat{A}^{\dagger}(\omega)]^{-1}\hat{A}^\dagger(\omega)=1$
has been inserted in the last line of Eq. \eqref{eq:alpha1},
which suggests to introduce an auxiliary
state $|Z(\omega)\rangle$ satisfying an equation
similar to Eq. \eqref{eq:rspw} but without the projector $\hat{Q}$
\begin{eqnarray}
\hat{A}(\omega)|Z(\omega)\rangle=\hat{z}|\Psi_0\rangle.\label{eq:rspw2}
\end{eqnarray}
Consequently, Eq. \eqref{eq:alpha1} can be recast into a symmetric form
\begin{eqnarray}
\langle \Psi_0|\hat{z}|\Psi(\omega)\rangle
&=&\langle Z(\omega)|\hat{A}^\dagger(\omega)|Z(\omega)\rangle\nonumber\\
&&+(\omega-i\gamma)|\langle Z(\omega)|\Psi_0\rangle|^2.\label{eq:azzSym}
\end{eqnarray}
Now the explicit dependence on the projector $\hat{Q}$, which makes
the design of a quantum algorithm difficult, has been removed
from both the response equation \eqref{eq:rspw2} and
the expression for the polarizability \eqref{eq:azzSym}.
Its effect is only reflected in the second term of Eq. \eqref{eq:azzSym}.
Likewise, the second part of $\alpha_{zz}(\omega)$ \eqref{eq:alpha} can be expressed in a similar
symmetric form
\begin{eqnarray}
\langle \Psi_0|\hat{z}|\Psi(-\omega)\rangle&=&
\langle Z(-\omega)|\hat{A}^\dagger(-\omega)|Z(-\omega)\rangle\nonumber\\
&&-(\omega-i\gamma)|\langle Z(-\omega)|\Psi_0\rangle|^2.\label{eq:azzSym2}
\end{eqnarray}
More explicitly, we can separate $\alpha_{zz}(\omega)$ into real and imaginary parts
\begin{eqnarray}
\alpha_{zz}(\omega)&=&\Re\alpha_{zz}(\omega)+i\Im\alpha_{zz}(\omega), \nonumber\\
\Re\alpha_{zz}(\omega)&=& \langle Z(\omega)|\hat{H}_0-E_0| Z(\omega)\rangle, \nonumber\\
&&+\langle Z(-\omega)|\hat{H}_0-E_0| Z(-\omega)\rangle
-\frac{\omega}{\gamma}\Im\alpha_{zz}(\omega), \nonumber\\
\Im\alpha_{zz}(\omega)&=& \gamma(\langle Z(\omega)|Z(\omega)\rangle
-\langle Z(-\omega)|Z(-\omega)\rangle\nonumber\\
&&
-|\langle Z(\omega)|\Psi_0\rangle|^2
+|\langle Z(-\omega)|\Psi_0\rangle|^2),\label{eq:ReIm}
\end{eqnarray}
where the expected symmetry relations $\Re\alpha_{zz}(-\omega)=\Re\alpha_{zz}(\omega)$ and
$\Im\alpha_{zz}(-\omega)=-\Im\alpha_{zz}(\omega)$ are obvious.
In fact, from Eq. \eqref{eq:rspw2} one can further find
$\langle\Psi_0|Z(\pm\omega)\rangle=\mp\frac{\langle\Psi_0|\hat{z}|\Psi_0\rangle}{\omega+i\eta}$,
such that the second terms in Eqs. \eqref{eq:azzSym} and \eqref{eq:azzSym2}
will cancel each other in $\alpha_{zz}(\omega)$.
Building upon the reformulation of the standard response theory,
Eqs. \eqref{eq:rspw2}-\eqref{eq:ReIm}, we are ready to present a quantum algorithm
for computing $\alpha_{zz}(\omega)$, using either
the quantum linear system algorithms\cite{harrow2009quantum,ambainis2010variable,
clader2013preconditioned,childs2017quantum,subacsi2019quantum} or the
variational hybrid quantum-classical algorithms\cite{xu2019variational,bravo2019variational}
for solving Eq. \eqref{eq:rspw2}.

{\it Quantum algorithm with an exponential speedup.}
We assume that the ground-state wavefunction $|\Psi_0\rangle$ and its associated energy $E_0$ have been
available either by QPE or VQE. The most challenging step for computing $\alpha_{zz}(\omega)$
is to solve the response equation \eqref{eq:rspw2},
which becomes a linear system of equation with dimension $D$ when
expressed in the full many-electron Hilbert space.
In this section, we prove that there is a quantum advantage for
computing $\alpha_{zz}(\omega)$ on quantum computers
over the classical FCI-based approaches\cite{olsen1985linear,koch1991analytical}
by using the HHL algorithm as a subroutine\cite{harrow2009quantum} to solve Eq. \eqref{eq:rspw2}.

Since $\hat{A}(\omega)$ \eqref{eq:Aw} in Eq. \eqref{eq:rspw2} is non-Hermitian for $\gamma>0$,
$|Z(\omega)\rangle$ can be determined using the HHL algorithm\cite{harrow2009quantum} either by
\begin{eqnarray}
\left[\begin{array}{cc}
0 & \hat{A}(\omega) \\
\hat{A}^\dagger(\omega) & 0 \\
\end{array}\right]
\left[\begin{array}{c}
0 \\
|Z(\omega)\rangle \\
\end{array}\right]
=
\left[\begin{array}{c}
\hat{z}|\Psi_0\rangle \\
0 \\
\end{array}\right],\label{eq:rspw3}
\end{eqnarray}
as suggested in the original work\cite{harrow2009quantum}
or by the following equivalent equation
\begin{eqnarray}
\hat{A}^\dagger(\omega)\hat{A}(\omega)|Z(\omega)\rangle=
\hat{A}^\dagger(\omega)\hat{z}|\Psi_0\rangle,\label{eq:rspw4}
\end{eqnarray}
with a Hermitian matrix on the left hand side (LHS), which has the same dimension
as Eq. \eqref{eq:rspw2} at the cost of increasing the condition number.
For a linear system of equations $\mathbf{Ax}=\mathbf{b}$, where $\mathbf{A}$ is a Hermitian
matrix of dimension $D$ with an eigendecomposition
$\mathbf{A}=\mathbf{U\Lambda U}^\dagger$,
the HHL algorithm\cite{harrow2009quantum} essentially prepares a solution
following the sequence $\mathbf{x}=\mathbf{U\Lambda^{-1}U^\dagger b}$.
The transformation to the eigenbasis of $\mathbf{A}$ and the backtransformation
are executed by QPE subroutines\cite{kitaev1995quantum}, which require the implementation of the controlled
time evolution $e^{i\mathbf{A}t}$,
while the realization of the nonunitary operation $\mathbf{\Lambda}^{-1}$ is through controlled
rotations also with the help of ancilla qubits.
The runtime complexity of the HHL algorithm is $O(\log(D)s^2\kappa^2/\epsilon)$\cite{harrow2009quantum},
where $\epsilon$ is the desired precision,
$s$ is the sparsity parameter specifying the maximal number of nonzero entries per row in $\mathbf{A}$,
and $\kappa$ is the condition number of $\mathbf{A}$, i.e.,
$\kappa=|\lambda _{\max}|/|\lambda _{\min}|$,
which is the ratio between the maximal and minimal eigenvalues by moduli of $\mathbf{A}$.
The real advantage of the HHL algorithm over classical algorithms crucially depends on the efficiency of
four major steps\cite{harrow2009quantum,aaronson2015read}:
(1) preparation of $\mathbf{b}$ on quantum computers,
(2) Hamiltonian simulation $e^{i\mathbf{A}t}$, (3) dependence of $\kappa$ on $D$,
and (4) readout of the output quantum state $|x\rangle=A^{-1}|b\rangle/\|A^{-1}|b\rangle\|$ encoding
the solution $\mathbf{x}$. Any slowdown in one of the steps could
kill the exponential speedup promised by the HHL algorithm.
Now we demonstrate that how an exponential speedup can be achieved
for computing $\alpha_{zz}(\omega)$ on quantum computers
by the following algorithm in a step-by-step way:

Step 1: Provided $|\Psi_0\rangle$ is available, the state $\hat{z}|\Psi_0\rangle$
on the right hand side (RHS) of Eq. \eqref{eq:rspw3} or $\hat{A}^\dagger(\omega)\hat{z}|\Psi_0\rangle$ in
Eq. \eqref{eq:rspw4} can be prepared with a cost of poly($N$) using the
linear combination of unitaries (LCU) algorithm\cite{Childs2012,berry2015simulating}.
This is because both the one-body perturbation $\hat{z}$ \eqref{eq:Vz}
and the Hamiltonian $\hat{H}_0$ \eqref{eq:H0} in $\hat{A}(\omega)$
can be expressed as a sum of poly($N$) Pauli matrices, e.g.,
\begin{eqnarray}
\hat{z}=\sum_{\mu}z_\mu P_\mu,\quad
P_\mu = \sigma_{\mu_1}\otimes\sigma_{\mu_2}\otimes\cdots \otimes\sigma_{\mu_K},\label{eq:VzP}
\end{eqnarray}
where $\sigma_{\mu_k}\in\{I_2,\sigma_x,\sigma_y,\sigma_z\}$ and the number of
terms is quadratic in $N$ for $\hat{z}$ \eqref{eq:Vz},
through a fermion-to-qubit mapping such as the Jordan-Wigner transformation\cite{jordan1928pauli}
or the Bravyi-Kitaev transformation\cite{bravyi2002fermionic,seeley2012bravyi}.

Step 2: Given the RHS of Eq. \eqref{eq:rspw3} (or Eq. \eqref{eq:rspw4}) prepared on quantum computers,
the HHL algorithm is applied to prepare a normalized solution state $|x\rangle=|Z(\omega)\rangle/\sqrt{\langle Z(\omega)|Z(\omega)\rangle}$ for Eq. \eqref{eq:rspw2}.
For molecular systems with $\hat{H}_0$ \eqref{eq:H0}, it is known
that the Hamiltonian simulation can be accomplished efficiently in poly($N$),
just as in applying QPE to the molecular ground state problem\cite{abrams1997simulation,
abrams1999quantum,aspuru2005simulated}. Because $\hat{H}_0$ involves at most two-body Coulomb interactions,
the sparsity parameter $s$ is only quartic in $N$.
Thus, the most crucial part for the runtime complexity of the HHL algorithm
is the condition number $\kappa$. On one hand, since $\hat{H}_0$ can be written as
a sum over $O(N^4)$ Pauli terms $\hat{H}_0=\sum_{\mu}h_{\mu}P_{\mu}$,
as for $\hat{z}$ in Eq. \eqref{eq:VzP},
$|\lambda _{\max}|$ of $\hat{H}_0$ (and $\hat{A}(\omega)$)
is bounded by a system-dependent constant $\max_\mu|h_{\mu}|$ times $N^4$.
Assuming we consider the scaling with respect to the variation of the system size
for systems of the same kind, such as water clusters of different sizes
in a given atomic orbital basis set, then $\max_\mu|h_{\mu}|$ is independent of $N$,
such that $|\lambda _{\max}|$ is of poly($N$).
On the other hand, the operator $\hat{H}_0-E_0-\omega$ becomes
singular whenever the frequency $\omega$ matches the excitation energy $\omega_{n0}$,
such that in the worst case $|\lambda _{\min}|$ of $\hat{A}^\dagger(\omega)\hat{A}(\omega)$ in Eq. \eqref{eq:rspw4}
is $\gamma^2$, and likewise for Eq. \eqref{eq:rspw3} $|\lambda _{\min}|$ equals $\gamma$.
In practice, the parameter $\gamma$ is a fixed input parameter for spectral resolution, which
determines the half width at half maximum (HWHM) of peaks in $\Im\alpha(\omega)$ \eqref{eq:ReIm}.
Thus, the condition numbers $\kappa$ for the coefficient matrices
in Eqs. \eqref{eq:rspw3} and \eqref{eq:rspw4} are polynomial in the system size $N$
instead of the dimension of the Hilbert space $D$, which is exponential in $N$.
This concludes that the runtime complexity of the HHL algorithm
for preparing the normalized solution state $|x\rangle$
from either Eq. \eqref{eq:rspw3} or \eqref{eq:rspw4}
is poly($N$).

Step 3: After applying the HHL algorithm to Eq. \eqref{eq:rspw3} or \eqref{eq:rspw4},
the first part of $\alpha_{zz}(\omega)$ \eqref{eq:alpha} can be computed from $|x\rangle$
using Eq. \eqref{eq:azzSym}
in poly($N$), without accessing its individual entry.
This is achieved by first noting that the norm of $|Z(\omega)\rangle$
required in Eq. \eqref{eq:azzSym} can
be computed using Eq. \eqref{eq:rspw2} as
\begin{eqnarray}
\langle Z(\omega)|Z(\omega)\rangle
=\langle\Psi_0|\hat{z}\hat{z}|\Psi_0\rangle/\langle x|\hat{A}^\dagger(\omega)\hat{A}(\omega)|x\rangle,\label{overlap}
\end{eqnarray}
which only requires the measurements of $\langle\Psi_0|\hat{z}\hat{z}|\Psi_0\rangle$ and $\langle x|\hat{A}^\dagger(\omega)\hat{A}(\omega)|x\rangle$.
Since the number of measurements is proportional to the number of terms\cite{wecker2015progress,mcclean2016theory}
in $\hat{z}\hat{z}$ and $\hat{A}^\dagger(\omega)\hat{A}(\omega)$, the cost
scales polynomially in $N$.
Then, the first term in Eq. \eqref{eq:azzSym} involving
$\langle Z(\omega)|\hat{H}_0|Z(\omega)\rangle$ can be obtained
from the measurement of $\langle x|\hat{H}_0|x\rangle=\sum_\mu h_\mu
\langle x|P_\mu|x\rangle$ in the same way as obtaining
the energy in VQE\cite{peruzzo2014variational,mcclean2016theory},
while the second term $|\langle Z(\omega)|\Psi_0\rangle|^2$
can be computed from $|\langle x|\Psi_0\rangle|^2$ by the SWAP test\cite{buhrman2001quantum,garcia2013swap}
or simply from $\frac{|\langle\Psi_0|\hat{z}|\Psi_0\rangle|^2}{\omega^2+\gamma^2}$ following
Eq. \eqref{eq:rspw2}. Therefore, the necessary information for computing $\alpha_{zz}(\omega)$ from the output state $|x\rangle$
of the HHL algorithm (and its counterpart for $|Z(-\omega)\rangle$)
can be accessed through $\langle x|\hat{H}_0^2|x\rangle$, $\langle x|\hat{H}_0|x\rangle$, and $|\langle x|\Psi_0\rangle|^2$ with a cost of poly($N$).

Using the above procedure, we show that the dynamic polarizability tensor $\alpha_{zz}(\omega)$ of molecules
can be computed on quantum computers with poly($N$) runtime complexity, achieving an exponential speedup compared
with the classical FCI-based approach\cite{olsen1985linear,koch1991analytical}.
This becomes possible due to the specialities of the molecular response problem,
such that all the limitations of the HHL algorithm can be overcome in
this application: the RHS of Eq. \eqref{eq:rspw2} can always be efficiently prepared given
$|\Psi_0\rangle$, $e^{i\hat{H}_0t}$ can be efficiently simulated
due to the sparse structure of $\hat{H}_0$ \eqref{eq:H0},
the condition numbers $\kappa$ for matrices in Eqs. \eqref{eq:rspw3} and \eqref{eq:rspw4} are polynomial in $N$,
and finally only partial information of the solution is required for computing
$\alpha_{zz}(\omega)$. Therefore, the molecular response problem is an ideal
application of the HHL algorithm\cite{harrow2009quantum}, and the same conclusion
is also generalizable to its improved
variants\cite{ambainis2010variable,clader2013preconditioned,childs2017quantum}.

{\it Variational hybrid quantum-classical algorithm.}
While the above HHL based quantum variant of our algorithm has a theoretically provable quantum advantage,
it is considered as a long-term algorithm in the sense that in general it requires a long circuit depth
and is less suitable for NISQ devices, even though there have been
recent experimental progresses on realizing the the HHL algorithm itself
on small scale problems\cite{cai2013experimental,pan2014experimental,barz2014two,zheng2017solving}.
To enable the computation of molecular response properties for potentially interesting larger molecules on near-term devices,
here we introduce a variational hybrid quantum-classical variant
by combining the same theoretical framework with the variational hybrid quantum-classical algorithms for linear systems of equations\cite{xu2019variational,bravo2019variational},
which like the VQE algorithm\cite{peruzzo2014variational,mcclean2016theory} have a much less requirement for circuit depth
and are more robust against noises due to the variational nature.

Specifically, suppose the ground state has been obtained by VQE through a variational parametrization
$|\Psi_0\rangle=U(\theta_0)|0\rangle$, where $U(\theta_0)$ represents
a parameterized unitary circuits with parameters $\theta_0$,
such as the unitary coupled cluster (UCC) ansatz\cite{peruzzo2014variational}
or the hardware efficient ansatz\cite{kandala2017hardware},
instead of solving Eq. \eqref{eq:rspw2} for $|Z(\omega)\rangle$ using the HHL algorithm,
we can design a parameterized ansatz for the
normalized state $|\theta\rangle=|Z(\omega)\rangle/\sqrt{\langle Z(\omega)|Z(\omega)\rangle}=U_Z(\theta)|0\rangle$.
Then, the solution of Eq. \eqref{eq:rspw2} can be found by minimizing the following cost function
\begin{eqnarray}
\mathcal{C}(\theta)=
\langle\Psi_0|\hat{z}\hat{z}|\Psi_0\rangle
\langle \theta|\hat{A}^\dagger(\omega)\hat{A}(\omega)|\theta\rangle
-|\langle\Psi_0|\hat{z}\hat{A}(\omega)|\theta\rangle|^2,\label{costFunctional}
\end{eqnarray}
with $\mathcal{C}(\theta)\ge 0$ due to the Cauchy-Schwarz inequality.
Since $\hat{A}(\omega)$ is always nonsingular for $\gamma>0$ regardless of $\omega$,
the solution of Eq. \eqref{eq:rspw2} is uniquely determined by the condition $\mathcal{C}_{\min}(\theta)=0$.
The two symmetric terms $\langle\Psi_0|\hat{z}\hat{z}|\Psi_0\rangle$ and $\langle \theta|\hat{A}^\dagger(\omega)\hat{A}(\omega)|\theta\rangle$
in Eq. \eqref{costFunctional} are exactly those appeared in Eq. \eqref{overlap}, and
hence can be evaluated in the same way through measurements after preparing
$|\Psi_0\rangle$ and $|\theta\rangle$, respectively. The overlap term
can be rewritten as $\langle\Psi_0|\hat{z}\hat{A}(\omega)|\theta\rangle=
\langle 0|U^\dagger (\theta_0)\hat{z}\hat{A}(\omega) U_Z(\theta)|0\rangle=
\sum_\mu \zeta_\mu \langle 0|U^\dagger(\theta_0) P_\mu U_Z(\theta)|0\rangle $,
where the expansions of $\hat{z}$ \eqref{eq:VzP} and $\hat{A}(\omega)$ were used to obtain
an expansion $\hat{z}\hat{A}(\omega)=\sum_\mu \zeta_\mu P_\mu$.
Terms like $\langle 0|U^\dagger(\theta_0) P_\mu U_Z(\theta)|0\rangle$ can be computed
in multiple ways, with the simplest choice being the standard Hadamard test.
It deserves to point out that improved techniques\cite{xu2019variational,bravo2019variational}
have been proposed for defining better cost functions 
and evaluating the overlap term with reduced requirements on the number of controlled
operations. Finally, once $|\theta\rangle$ has been determined variationally,
$\alpha_{zz}(\omega)$ \eqref{eq:ReIm} can be computed in exactly the same way following
Step 3 in the previous section. In practice, errors in computing
$\mathcal{C}(\theta)$ and $\alpha_{zz}(\omega)$ due to noises can be
mitigated using the available techniques\cite{li2017efficient,temme2017error,mcardle2019error,kandala2019error,torlai2020precise}
developed for VQE to achieve better accuracy. Thus, together with the VQE algorithm\cite{peruzzo2014variational,mcclean2016theory}
for the ground state $|\Psi_0\rangle$, this variational hybrid quantum-classical variant provides a more
practical way for computing molecular response properties on near-term devices.

{\it Extensions to general response properties.}
Both two variants of the proposed algorithm can be generalized to compute general
linear and nonlinear response properties,
using the same idea of first deriving a symmetric expression for the target property
by introducing appropriate auxiliary states, and then determining
these states by solving response equations with appropriate quantum or hybrid algorithms.
For off-diagonal components of $\alpha_{ij}(\omega)$, e.g., $\alpha_{xz}(\omega)$
containing a form of $\langle \Psi_0|\hat{x}\hat{Q}[\hat{Q}\hat{A}(\omega)\hat{Q}]^{-1}\hat{Q}\hat{z}|\Psi_0\rangle$,
a symmetric expression can be derived by applying the polarization identity,
which involves a linear combination of four symmetric terms $\langle \Psi_0|(\hat{x}\pm(i)\hat{z})^\dagger
\hat{Q}[\hat{Q}\hat{A}(\omega)\hat{Q}]^{-1}\hat{Q}(\hat{x}\pm(i)\hat{z})|\Psi_0\rangle$, viz.,
\begin{eqnarray}
&&\Re\langle \Psi_0|\hat{x}\hat{Q}[\hat{Q}\hat{A}(\omega)\hat{Q}]^{-1}\hat{Q}\hat{z}|\Psi_0\rangle\nonumber\\
&=&
\frac{1}{4}(\langle \Psi_0|(\hat{x}+\hat{z})^\dagger
\hat{Q}[\hat{Q}\hat{A}(\omega)\hat{Q}]^{-1}\hat{Q}(\hat{x}+\hat{z})|\Psi_0\rangle \nonumber\\
&&-\langle \Psi_0|(\hat{x}-\hat{z})^\dagger
\hat{Q}[\hat{Q}\hat{A}(\omega)\hat{Q}]^{-1}\hat{Q}(\hat{x}-\hat{z})|\Psi_0\rangle),
\end{eqnarray}
and
\begin{eqnarray}
&&\Im\langle \Psi_0|\hat{x}\hat{Q}[\hat{Q}\hat{A}(\omega)\hat{Q}]^{-1}\hat{Q}\hat{z}|\Psi_0\rangle\nonumber\\
&=&
-\frac{1}{4}(\langle \Psi_0|(\hat{x}+i\hat{z})^\dagger
\hat{Q}[\hat{Q}\hat{A}(\omega)\hat{Q}]^{-1}\hat{Q}(\hat{x}+i\hat{z})|\Psi_0\rangle \nonumber\\
&&-\langle \Psi_0|(\hat{x}-i\hat{z})^\dagger
\hat{Q}[\hat{Q}\hat{A}(\omega)\hat{Q}]^{-1}\hat{Q}(\hat{x}-i\hat{z})|\Psi_0\rangle).
\end{eqnarray}
Each of them can be computed using the same algorithm for $\alpha_{zz}(\omega)$.
As an important example for nonlinear response functions, we consider the resonant
inelastic X-ray scattering (RIXS) amplitudes\cite{ament2011resonant,norman2018simulating}
for probing elementary excitations in complex correlated electron systems.
It is given by the Kramers-Heisenberg formula\cite{kramers1925streuung}
\begin{eqnarray}
\mathcal{F}^{f0}_{zz}(\omega)
&=&
\sum_{n}\left[
\frac{\langle \Psi_f|\hat{z}|\Psi_n\rangle\langle\Psi_n|\hat{z}|\Psi_0\rangle}
{\omega_{n0}-(\omega+i\gamma)}\right.\nonumber\\
&&\left.+\frac{\langle \Psi_f|\hat{z}|\Psi_n\rangle\langle \Psi_n|\hat{z}|\Psi_0\rangle}
{\omega_{n0}+(\omega'+i\gamma)}\right],\label{eq:SOS2}
\end{eqnarray}
where $\omega'\triangleq\omega-\omega_{f0}$ and $|\Psi_f\rangle$
represents the final state of interest involved in the inelastic scattering process.
Since Eq. \eqref{eq:SOS2} takes a similar form as Eq. \eqref{eq:SOS} for $\alpha_{zz}(\omega)$,
a similar strategy can be designed to compute
the scattering cross section $|\mathcal{F}^{f0}_{zz}(\omega)|^2$
\cite{ament2011resonant}.
In particular, within the rotating wave approximation,
where the second term of Eq. \eqref{eq:SOS2} is neglected,
the scattering cross section is simply given by
$|\mathcal{F}^{f0}_{zz}(\omega)|^2=|\langle\Psi_f|\hat{z}|Z(\omega)\rangle|^2$
with the same $|Z(\omega)\rangle$ in Eq. \eqref{eq:rspw2},
which can be computed by modifications of the SWAP test\cite{buhrman2001quantum}

In summary, we presented a general algorithm for computing molecular
response properties on quantum computers.
The most demanding step involves a set of linear systems of equations for auxiliary quantum states,
which can be solved either by quantum algorithms\cite{harrow2009quantum,ambainis2010variable,
clader2013preconditioned,childs2017quantum,subacsi2019quantum} or variational hybrid quantum-classical algorithms\cite{xu2019variational,bravo2019variational}.
The resulting two variants enable the computation of molecular
response properties for interested frequencies directly.
While the later variational hybrid variant is more suitable for near-term applications,
we showed that the former with the HHL algorithm\cite{harrow2009quantum} as a subroutine
has a provable quantum speedup over existing classical
FCI-based approach\cite{olsen1985linear,koch1991analytical}.
Our work provides a new theoretical evidence that
quantum chemistry is a promising area that will benefit from quantum computation.
Enabling accurate and efficient predictions of molecular response properties on quantum
computers will potentially open up a broad range of new applications of quantum computation in material
science and drug discovery in the near future.

{\it Acknowledgements.} The author (Z.L.) acknowledges Garnet Kin-Lic Chan, Jiajun Ren, Man-Hong Yung, and Zheng Li for critically reading the manuscript and helpful comments. We thank Zi-Yong Ge, Zhengan Wang, and Kai Xu for discussions.
This work was supported by the National Natural Science Foundation of China (Grants
No. 21973003) and the Beijing Normal University Startup Package.

{\it Note: During the review process of this work, we became
aware of a related work in Ref. \cite{PhysRevResearch.2.033043},
which proposed a different algorithm for constructing
the linear response functions of molecules via
quantum phase estimation and statistical sampling.
}

\bibliographystyle{apsrev4-1}
\bibliography{references}

\begin{thebibliography}{66}%
\makeatletter
\providecommand \@ifxundefined [1]{%
 \@ifx{#1\undefined}
}%
\providecommand \@ifnum [1]{%
 \ifnum #1\expandafter \@firstoftwo
 \else \expandafter \@secondoftwo
 \fi
}%
\providecommand \@ifx [1]{%
 \ifx #1\expandafter \@firstoftwo
 \else \expandafter \@secondoftwo
 \fi
}%
\providecommand \natexlab [1]{#1}%
\providecommand \enquote  [1]{``#1''}%
\providecommand \bibnamefont  [1]{#1}%
\providecommand \bibfnamefont [1]{#1}%
\providecommand \citenamefont [1]{#1}%
\providecommand \href@noop [0]{\@secondoftwo}%
\providecommand \href [0]{\begingroup \@sanitize@url \@href}%
\providecommand \@href[1]{\@@startlink{#1}\@@href}%
\providecommand \@@href[1]{\endgroup#1\@@endlink}%
\providecommand \@sanitize@url [0]{\catcode `\\12\catcode `\$12\catcode
  `\&12\catcode `\#12\catcode `\^12\catcode `\_12\catcode `\%12\relax}%
\providecommand \@@startlink[1]{}%
\providecommand \@@endlink[0]{}%
\providecommand \url  [0]{\begingroup\@sanitize@url \@url }%
\providecommand \@url [1]{\endgroup\@href {#1}{\urlprefix }}%
\providecommand \urlprefix  [0]{URL }%
\providecommand \Eprint [0]{\href }%
\providecommand \doibase [0]{http://dx.doi.org/}%
\providecommand \selectlanguage [0]{\@gobble}%
\providecommand \bibinfo  [0]{\@secondoftwo}%
\providecommand \bibfield  [0]{\@secondoftwo}%
\providecommand \translation [1]{[#1]}%
\providecommand \BibitemOpen [0]{}%
\providecommand \bibitemStop [0]{}%
\providecommand \bibitemNoStop [0]{.\EOS\space}%
\providecommand \EOS [0]{\spacefactor3000\relax}%
\providecommand \BibitemShut  [1]{\csname bibitem#1\endcsname}%
\let\auto@bib@innerbib\@empty
\bibitem [{\citenamefont {Boyd}(2003)}]{boyd2003nonlinear}%
  \BibitemOpen
  \bibfield  {author} {\bibinfo {author} {\bibfnamefont {R.~W.}\ \bibnamefont
  {Boyd}},\ }\href@noop {} {\emph {\bibinfo {title} {Nonlinear optics}}}\
  (\bibinfo  {publisher} {Elsevier},\ \bibinfo {year} {2003})\BibitemShut
  {NoStop}%
\bibitem [{\citenamefont {Norman}\ \emph {et~al.}(2018)\citenamefont {Norman},
  \citenamefont {Ruud},\ and\ \citenamefont {Saue}}]{norman2018principles}%
  \BibitemOpen
  \bibfield  {author} {\bibinfo {author} {\bibfnamefont {P.}~\bibnamefont
  {Norman}}, \bibinfo {author} {\bibfnamefont {K.}~\bibnamefont {Ruud}}, \ and\
  \bibinfo {author} {\bibfnamefont {T.}~\bibnamefont {Saue}},\ }\href@noop {}
  {\emph {\bibinfo {title} {Principles and practices of molecular properties:
  Theory, modeling, and simulations}}}\ (\bibinfo  {publisher} {John Wiley \&
  Sons},\ \bibinfo {year} {2018})\BibitemShut {NoStop}%
\bibitem [{\citenamefont {Casimir}\ and\ \citenamefont
  {Polder}(1948)}]{casimir1948influence}%
  \BibitemOpen
  \bibfield  {author} {\bibinfo {author} {\bibfnamefont {H.~B.}\ \bibnamefont
  {Casimir}}\ and\ \bibinfo {author} {\bibfnamefont {D.}~\bibnamefont
  {Polder}},\ }\href@noop {} {\bibfield  {journal} {\bibinfo  {journal}
  {Physical Review}\ }\textbf {\bibinfo {volume} {73}},\ \bibinfo {pages} {360}
  (\bibinfo {year} {1948})}\BibitemShut {NoStop}%
\bibitem [{\citenamefont {Helgaker}\ \emph {et~al.}(2012)\citenamefont
  {Helgaker}, \citenamefont {Coriani}, \citenamefont {J{\o}rgensen},
  \citenamefont {Kristensen}, \citenamefont {Olsen},\ and\ \citenamefont
  {Ruud}}]{helgaker2012recent}%
  \BibitemOpen
  \bibfield  {author} {\bibinfo {author} {\bibfnamefont {T.}~\bibnamefont
  {Helgaker}}, \bibinfo {author} {\bibfnamefont {S.}~\bibnamefont {Coriani}},
  \bibinfo {author} {\bibfnamefont {P.}~\bibnamefont {J{\o}rgensen}}, \bibinfo
  {author} {\bibfnamefont {K.}~\bibnamefont {Kristensen}}, \bibinfo {author}
  {\bibfnamefont {J.}~\bibnamefont {Olsen}}, \ and\ \bibinfo {author}
  {\bibfnamefont {K.}~\bibnamefont {Ruud}},\ }\href@noop {} {\bibfield
  {journal} {\bibinfo  {journal} {Chemical Reviews}\ }\textbf {\bibinfo
  {volume} {112}},\ \bibinfo {pages} {543} (\bibinfo {year}
  {2012})}\BibitemShut {NoStop}%
\bibitem [{\citenamefont {Norman}\ and\ \citenamefont
  {Dreuw}(2018)}]{norman2018simulating}%
  \BibitemOpen
  \bibfield  {author} {\bibinfo {author} {\bibfnamefont {P.}~\bibnamefont
  {Norman}}\ and\ \bibinfo {author} {\bibfnamefont {A.}~\bibnamefont {Dreuw}},\
  }\href@noop {} {\bibfield  {journal} {\bibinfo  {journal} {Chemical Reviews}\
  }\textbf {\bibinfo {volume} {118}},\ \bibinfo {pages} {7208} (\bibinfo {year}
  {2018})}\BibitemShut {NoStop}%
\bibitem [{\citenamefont {Szabo}\ and\ \citenamefont
  {Ostlund}(2012)}]{szabo2012modern}%
  \BibitemOpen
  \bibfield  {author} {\bibinfo {author} {\bibfnamefont {A.}~\bibnamefont
  {Szabo}}\ and\ \bibinfo {author} {\bibfnamefont {N.~S.}\ \bibnamefont
  {Ostlund}},\ }\href@noop {} {\emph {\bibinfo {title} {Modern quantum
  chemistry: introduction to advanced electronic structure theory}}}\ (\bibinfo
   {publisher} {Courier Corporation},\ \bibinfo {year} {2012})\BibitemShut
  {NoStop}%
\bibitem [{\citenamefont {Olsen}\ and\ \citenamefont
  {J{\o}rgensen}(1985)}]{olsen1985linear}%
  \BibitemOpen
  \bibfield  {author} {\bibinfo {author} {\bibfnamefont {J.}~\bibnamefont
  {Olsen}}\ and\ \bibinfo {author} {\bibfnamefont {P.}~\bibnamefont
  {J{\o}rgensen}},\ }\href@noop {} {\bibfield  {journal} {\bibinfo  {journal}
  {The Journal of Chemical Physics}\ }\textbf {\bibinfo {volume} {82}},\
  \bibinfo {pages} {3235} (\bibinfo {year} {1985})}\BibitemShut {NoStop}%
\bibitem [{\citenamefont {Koch}\ and\ \citenamefont
  {Harrison}(1991)}]{koch1991analytical}%
  \BibitemOpen
  \bibfield  {author} {\bibinfo {author} {\bibfnamefont {H.}~\bibnamefont
  {Koch}}\ and\ \bibinfo {author} {\bibfnamefont {R.~J.}\ \bibnamefont
  {Harrison}},\ }\href@noop {} {\bibfield  {journal} {\bibinfo  {journal} {The
  Journal of Chemical Physics}\ }\textbf {\bibinfo {volume} {95}},\ \bibinfo
  {pages} {7479} (\bibinfo {year} {1991})}\BibitemShut {NoStop}%
\bibitem [{\citenamefont {Cohen}\ \emph {et~al.}(2008)\citenamefont {Cohen},
  \citenamefont {Mori-S{\'a}nchez},\ and\ \citenamefont
  {Yang}}]{cohen2008insights}%
  \BibitemOpen
  \bibfield  {author} {\bibinfo {author} {\bibfnamefont {A.~J.}\ \bibnamefont
  {Cohen}}, \bibinfo {author} {\bibfnamefont {P.}~\bibnamefont
  {Mori-S{\'a}nchez}}, \ and\ \bibinfo {author} {\bibfnamefont
  {W.}~\bibnamefont {Yang}},\ }\href@noop {} {\bibfield  {journal} {\bibinfo
  {journal} {Science}\ }\textbf {\bibinfo {volume} {321}},\ \bibinfo {pages}
  {792} (\bibinfo {year} {2008})}\BibitemShut {NoStop}%
\bibitem [{\citenamefont {Kent}\ and\ \citenamefont
  {Kotliar}(2018)}]{kent2018toward}%
  \BibitemOpen
  \bibfield  {author} {\bibinfo {author} {\bibfnamefont {P.~R.}\ \bibnamefont
  {Kent}}\ and\ \bibinfo {author} {\bibfnamefont {G.}~\bibnamefont {Kotliar}},\
  }\href@noop {} {\bibfield  {journal} {\bibinfo  {journal} {Science}\ }\textbf
  {\bibinfo {volume} {361}},\ \bibinfo {pages} {348} (\bibinfo {year}
  {2018})}\BibitemShut {NoStop}%
\bibitem [{\citenamefont {Feynman}(1982)}]{feynman1982simulating}%
  \BibitemOpen
  \bibfield  {author} {\bibinfo {author} {\bibfnamefont {R.~P.}\ \bibnamefont
  {Feynman}},\ }\href@noop {} {\bibfield  {journal} {\bibinfo  {journal}
  {International Journal of Theoretical Physics}\ }\textbf {\bibinfo {volume}
  {21}},\ \bibinfo {pages} {467} (\bibinfo {year} {1982})}\BibitemShut
  {NoStop}%
\bibitem [{\citenamefont {Lloyd}(1996)}]{lloyd1996universal}%
  \BibitemOpen
  \bibfield  {author} {\bibinfo {author} {\bibfnamefont {S.}~\bibnamefont
  {Lloyd}},\ }\href@noop {} {\bibfield  {journal} {\bibinfo  {journal}
  {Science}\ }\textbf {\bibinfo {volume} {273}},\ \bibinfo {pages} {1073}
  (\bibinfo {year} {1996})}\BibitemShut {NoStop}%
\bibitem [{\citenamefont {Abrams}\ and\ \citenamefont
  {Lloyd}(1997)}]{abrams1997simulation}%
  \BibitemOpen
  \bibfield  {author} {\bibinfo {author} {\bibfnamefont {D.~S.}\ \bibnamefont
  {Abrams}}\ and\ \bibinfo {author} {\bibfnamefont {S.}~\bibnamefont {Lloyd}},\
  }\href@noop {} {\bibfield  {journal} {\bibinfo  {journal} {Physical Review
  Letters}\ }\textbf {\bibinfo {volume} {79}},\ \bibinfo {pages} {2586}
  (\bibinfo {year} {1997})}\BibitemShut {NoStop}%
\bibitem [{\citenamefont {Abrams}\ and\ \citenamefont
  {Lloyd}(1999)}]{abrams1999quantum}%
  \BibitemOpen
  \bibfield  {author} {\bibinfo {author} {\bibfnamefont {D.~S.}\ \bibnamefont
  {Abrams}}\ and\ \bibinfo {author} {\bibfnamefont {S.}~\bibnamefont {Lloyd}},\
  }\href@noop {} {\bibfield  {journal} {\bibinfo  {journal} {Physical Review
  Letters}\ }\textbf {\bibinfo {volume} {83}},\ \bibinfo {pages} {5162}
  (\bibinfo {year} {1999})}\BibitemShut {NoStop}%
\bibitem [{\citenamefont {Ortiz}\ \emph {et~al.}(2001)\citenamefont {Ortiz},
  \citenamefont {Gubernatis}, \citenamefont {Knill},\ and\ \citenamefont
  {Laflamme}}]{ortiz2001quantum}%
  \BibitemOpen
  \bibfield  {author} {\bibinfo {author} {\bibfnamefont {G.}~\bibnamefont
  {Ortiz}}, \bibinfo {author} {\bibfnamefont {J.~E.}\ \bibnamefont
  {Gubernatis}}, \bibinfo {author} {\bibfnamefont {E.}~\bibnamefont {Knill}}, \
  and\ \bibinfo {author} {\bibfnamefont {R.}~\bibnamefont {Laflamme}},\
  }\href@noop {} {\bibfield  {journal} {\bibinfo  {journal} {Physical Review
  A}\ }\textbf {\bibinfo {volume} {64}},\ \bibinfo {pages} {022319} (\bibinfo
  {year} {2001})}\BibitemShut {NoStop}%
\bibitem [{\citenamefont {Somma}\ \emph {et~al.}(2002)\citenamefont {Somma},
  \citenamefont {Ortiz}, \citenamefont {Gubernatis}, \citenamefont {Knill},\
  and\ \citenamefont {Laflamme}}]{somma2002simulating}%
  \BibitemOpen
  \bibfield  {author} {\bibinfo {author} {\bibfnamefont {R.}~\bibnamefont
  {Somma}}, \bibinfo {author} {\bibfnamefont {G.}~\bibnamefont {Ortiz}},
  \bibinfo {author} {\bibfnamefont {J.~E.}\ \bibnamefont {Gubernatis}},
  \bibinfo {author} {\bibfnamefont {E.}~\bibnamefont {Knill}}, \ and\ \bibinfo
  {author} {\bibfnamefont {R.}~\bibnamefont {Laflamme}},\ }\href@noop {}
  {\bibfield  {journal} {\bibinfo  {journal} {Physical Review A}\ }\textbf
  {\bibinfo {volume} {65}},\ \bibinfo {pages} {042323} (\bibinfo {year}
  {2002})}\BibitemShut {NoStop}%
\bibitem [{\citenamefont {Georgescu}\ \emph {et~al.}(2014)\citenamefont
  {Georgescu}, \citenamefont {Ashhab},\ and\ \citenamefont
  {Nori}}]{georgescu2014quantum}%
  \BibitemOpen
  \bibfield  {author} {\bibinfo {author} {\bibfnamefont {I.~M.}\ \bibnamefont
  {Georgescu}}, \bibinfo {author} {\bibfnamefont {S.}~\bibnamefont {Ashhab}}, \
  and\ \bibinfo {author} {\bibfnamefont {F.}~\bibnamefont {Nori}},\ }\href@noop
  {} {\bibfield  {journal} {\bibinfo  {journal} {Reviews of Modern Physics}\
  }\textbf {\bibinfo {volume} {86}},\ \bibinfo {pages} {153} (\bibinfo {year}
  {2014})}\BibitemShut {NoStop}%
\bibitem [{\citenamefont {Cao}\ \emph {et~al.}(2019)\citenamefont {Cao},
  \citenamefont {Romero}, \citenamefont {Olson}, \citenamefont {Degroote},
  \citenamefont {Johnson}, \citenamefont {Kieferov{\'a}}, \citenamefont
  {Kivlichan}, \citenamefont {Menke}, \citenamefont {Peropadre}, \citenamefont
  {Sawaya} \emph {et~al.}}]{cao2019quantum}%
  \BibitemOpen
  \bibfield  {author} {\bibinfo {author} {\bibfnamefont {Y.}~\bibnamefont
  {Cao}}, \bibinfo {author} {\bibfnamefont {J.}~\bibnamefont {Romero}},
  \bibinfo {author} {\bibfnamefont {J.~P.}\ \bibnamefont {Olson}}, \bibinfo
  {author} {\bibfnamefont {M.}~\bibnamefont {Degroote}}, \bibinfo {author}
  {\bibfnamefont {P.~D.}\ \bibnamefont {Johnson}}, \bibinfo {author}
  {\bibfnamefont {M.}~\bibnamefont {Kieferov{\'a}}}, \bibinfo {author}
  {\bibfnamefont {I.~D.}\ \bibnamefont {Kivlichan}}, \bibinfo {author}
  {\bibfnamefont {T.}~\bibnamefont {Menke}}, \bibinfo {author} {\bibfnamefont
  {B.}~\bibnamefont {Peropadre}}, \bibinfo {author} {\bibfnamefont {N.~P.}\
  \bibnamefont {Sawaya}},  \emph {et~al.},\ }\href@noop {} {\bibfield
  {journal} {\bibinfo  {journal} {Chemical Reviews}\ }\textbf {\bibinfo
  {volume} {119}},\ \bibinfo {pages} {10856} (\bibinfo {year}
  {2019})}\BibitemShut {NoStop}%
\bibitem [{\citenamefont {McArdle}\ \emph {et~al.}(2020)\citenamefont
  {McArdle}, \citenamefont {Endo}, \citenamefont {Aspuru-Guzik}, \citenamefont
  {Benjamin},\ and\ \citenamefont {Yuan}}]{yuan2020}%
  \BibitemOpen
  \bibfield  {author} {\bibinfo {author} {\bibfnamefont {S.}~\bibnamefont
  {McArdle}}, \bibinfo {author} {\bibfnamefont {S.}~\bibnamefont {Endo}},
  \bibinfo {author} {\bibfnamefont {A.}~\bibnamefont {Aspuru-Guzik}}, \bibinfo
  {author} {\bibfnamefont {S.~C.}\ \bibnamefont {Benjamin}}, \ and\ \bibinfo
  {author} {\bibfnamefont {X.}~\bibnamefont {Yuan}},\ }\href@noop {} {\bibfield
   {journal} {\bibinfo  {journal} {Reviews of Modern Physics}\ }\textbf
  {\bibinfo {volume} {92}},\ \bibinfo {pages} {015003} (\bibinfo {year}
  {2020})}\BibitemShut {NoStop}%
\bibitem [{\citenamefont {Kitaev}(1995)}]{kitaev1995quantum}%
  \BibitemOpen
  \bibfield  {author} {\bibinfo {author} {\bibfnamefont {A.~Y.}\ \bibnamefont
  {Kitaev}},\ }\href@noop {} {\bibfield  {journal} {\bibinfo  {journal} {arXiv
  preprint quant-ph/9511026}\ } (\bibinfo {year} {1995})}\BibitemShut {NoStop}%
\bibitem [{\citenamefont {Aspuru-Guzik}\ \emph {et~al.}(2005)\citenamefont
  {Aspuru-Guzik}, \citenamefont {Dutoi}, \citenamefont {Love},\ and\
  \citenamefont {Head-Gordon}}]{aspuru2005simulated}%
  \BibitemOpen
  \bibfield  {author} {\bibinfo {author} {\bibfnamefont {A.}~\bibnamefont
  {Aspuru-Guzik}}, \bibinfo {author} {\bibfnamefont {A.~D.}\ \bibnamefont
  {Dutoi}}, \bibinfo {author} {\bibfnamefont {P.~J.}\ \bibnamefont {Love}}, \
  and\ \bibinfo {author} {\bibfnamefont {M.}~\bibnamefont {Head-Gordon}},\
  }\href@noop {} {\bibfield  {journal} {\bibinfo  {journal} {Science}\ }\textbf
  {\bibinfo {volume} {309}},\ \bibinfo {pages} {1704} (\bibinfo {year}
  {2005})}\BibitemShut {NoStop}%
\bibitem [{\citenamefont {Kassal}\ and\ \citenamefont
  {Aspuru-Guzik}(2009)}]{kassal2009quantum}%
  \BibitemOpen
  \bibfield  {author} {\bibinfo {author} {\bibfnamefont {I.}~\bibnamefont
  {Kassal}}\ and\ \bibinfo {author} {\bibfnamefont {A.}~\bibnamefont
  {Aspuru-Guzik}},\ }\href@noop {} {\bibfield  {journal} {\bibinfo  {journal}
  {The Journal of Chemical Physics}\ }\textbf {\bibinfo {volume} {131}},\
  \bibinfo {pages} {224102} (\bibinfo {year} {2009})}\BibitemShut {NoStop}%
\bibitem [{\citenamefont {O'Brien}\ \emph {et~al.}(2019)\citenamefont
  {O'Brien}, \citenamefont {Senjean}, \citenamefont {Sagastizabal},
  \citenamefont {Bonet-Monroig}, \citenamefont {Dutkiewicz}, \citenamefont
  {Buda}, \citenamefont {DiCarlo},\ and\ \citenamefont
  {Visscher}}]{o2019calculating}%
  \BibitemOpen
  \bibfield  {author} {\bibinfo {author} {\bibfnamefont {T.~E.}\ \bibnamefont
  {O'Brien}}, \bibinfo {author} {\bibfnamefont {B.}~\bibnamefont {Senjean}},
  \bibinfo {author} {\bibfnamefont {R.}~\bibnamefont {Sagastizabal}}, \bibinfo
  {author} {\bibfnamefont {X.}~\bibnamefont {Bonet-Monroig}}, \bibinfo {author}
  {\bibfnamefont {A.}~\bibnamefont {Dutkiewicz}}, \bibinfo {author}
  {\bibfnamefont {F.}~\bibnamefont {Buda}}, \bibinfo {author} {\bibfnamefont
  {L.}~\bibnamefont {DiCarlo}}, \ and\ \bibinfo {author} {\bibfnamefont
  {L.}~\bibnamefont {Visscher}},\ }\href@noop {} {\bibfield  {journal}
  {\bibinfo  {journal} {npj Quantum Information}\ }\textbf {\bibinfo {volume}
  {5}},\ \bibinfo {pages} {1} (\bibinfo {year} {2019})}\BibitemShut {NoStop}%
\bibitem [{\citenamefont {Lanyon}\ \emph {et~al.}(2010)\citenamefont {Lanyon},
  \citenamefont {Whitfield}, \citenamefont {Gillett}, \citenamefont {Goggin},
  \citenamefont {Almeida}, \citenamefont {Kassal}, \citenamefont {Biamonte},
  \citenamefont {Mohseni}, \citenamefont {Powell}, \citenamefont {Barbieri}
  \emph {et~al.}}]{lanyon2010towards}%
  \BibitemOpen
  \bibfield  {author} {\bibinfo {author} {\bibfnamefont {B.~P.}\ \bibnamefont
  {Lanyon}}, \bibinfo {author} {\bibfnamefont {J.~D.}\ \bibnamefont
  {Whitfield}}, \bibinfo {author} {\bibfnamefont {G.~G.}\ \bibnamefont
  {Gillett}}, \bibinfo {author} {\bibfnamefont {M.~E.}\ \bibnamefont {Goggin}},
  \bibinfo {author} {\bibfnamefont {M.~P.}\ \bibnamefont {Almeida}}, \bibinfo
  {author} {\bibfnamefont {I.}~\bibnamefont {Kassal}}, \bibinfo {author}
  {\bibfnamefont {J.~D.}\ \bibnamefont {Biamonte}}, \bibinfo {author}
  {\bibfnamefont {M.}~\bibnamefont {Mohseni}}, \bibinfo {author} {\bibfnamefont
  {B.~J.}\ \bibnamefont {Powell}}, \bibinfo {author} {\bibfnamefont
  {M.}~\bibnamefont {Barbieri}},  \emph {et~al.},\ }\href@noop {} {\bibfield
  {journal} {\bibinfo  {journal} {Nature Chemistry}\ }\textbf {\bibinfo
  {volume} {2}},\ \bibinfo {pages} {106} (\bibinfo {year} {2010})}\BibitemShut
  {NoStop}%
\bibitem [{\citenamefont {Du}\ \emph {et~al.}(2010)\citenamefont {Du},
  \citenamefont {Xu}, \citenamefont {Peng}, \citenamefont {Wang}, \citenamefont
  {Wu},\ and\ \citenamefont {Lu}}]{du2010nmr}%
  \BibitemOpen
  \bibfield  {author} {\bibinfo {author} {\bibfnamefont {J.}~\bibnamefont
  {Du}}, \bibinfo {author} {\bibfnamefont {N.}~\bibnamefont {Xu}}, \bibinfo
  {author} {\bibfnamefont {X.}~\bibnamefont {Peng}}, \bibinfo {author}
  {\bibfnamefont {P.}~\bibnamefont {Wang}}, \bibinfo {author} {\bibfnamefont
  {S.}~\bibnamefont {Wu}}, \ and\ \bibinfo {author} {\bibfnamefont
  {D.}~\bibnamefont {Lu}},\ }\href@noop {} {\bibfield  {journal} {\bibinfo
  {journal} {Physical Review Letters}\ }\textbf {\bibinfo {volume} {104}},\
  \bibinfo {pages} {030502} (\bibinfo {year} {2010})}\BibitemShut {NoStop}%
\bibitem [{\citenamefont {Wang}\ \emph {et~al.}(2015)\citenamefont {Wang},
  \citenamefont {Dolde}, \citenamefont {Biamonte}, \citenamefont {Babbush},
  \citenamefont {Bergholm}, \citenamefont {Yang}, \citenamefont {Jakobi},
  \citenamefont {Neumann}, \citenamefont {Aspuru-Guzik}, \citenamefont
  {Whitfield} \emph {et~al.}}]{wang2015quantum}%
  \BibitemOpen
  \bibfield  {author} {\bibinfo {author} {\bibfnamefont {Y.}~\bibnamefont
  {Wang}}, \bibinfo {author} {\bibfnamefont {F.}~\bibnamefont {Dolde}},
  \bibinfo {author} {\bibfnamefont {J.}~\bibnamefont {Biamonte}}, \bibinfo
  {author} {\bibfnamefont {R.}~\bibnamefont {Babbush}}, \bibinfo {author}
  {\bibfnamefont {V.}~\bibnamefont {Bergholm}}, \bibinfo {author}
  {\bibfnamefont {S.}~\bibnamefont {Yang}}, \bibinfo {author} {\bibfnamefont
  {I.}~\bibnamefont {Jakobi}}, \bibinfo {author} {\bibfnamefont
  {P.}~\bibnamefont {Neumann}}, \bibinfo {author} {\bibfnamefont
  {A.}~\bibnamefont {Aspuru-Guzik}}, \bibinfo {author} {\bibfnamefont {J.~D.}\
  \bibnamefont {Whitfield}},  \emph {et~al.},\ }\href@noop {} {\bibfield
  {journal} {\bibinfo  {journal} {ACS Nano}\ }\textbf {\bibinfo {volume} {9}},\
  \bibinfo {pages} {7769} (\bibinfo {year} {2015})}\BibitemShut {NoStop}%
\bibitem [{\citenamefont {O'Malley}\ \emph {et~al.}(2016)\citenamefont
  {O'Malley}, \citenamefont {Babbush}, \citenamefont {Kivlichan}, \citenamefont
  {Romero}, \citenamefont {McClean}, \citenamefont {Barends}, \citenamefont
  {Kelly}, \citenamefont {Roushan}, \citenamefont {Tranter}, \citenamefont
  {Ding} \emph {et~al.}}]{o2016scalable}%
  \BibitemOpen
  \bibfield  {author} {\bibinfo {author} {\bibfnamefont {P.~J.}\ \bibnamefont
  {O'Malley}}, \bibinfo {author} {\bibfnamefont {R.}~\bibnamefont {Babbush}},
  \bibinfo {author} {\bibfnamefont {I.~D.}\ \bibnamefont {Kivlichan}}, \bibinfo
  {author} {\bibfnamefont {J.}~\bibnamefont {Romero}}, \bibinfo {author}
  {\bibfnamefont {J.~R.}\ \bibnamefont {McClean}}, \bibinfo {author}
  {\bibfnamefont {R.}~\bibnamefont {Barends}}, \bibinfo {author} {\bibfnamefont
  {J.}~\bibnamefont {Kelly}}, \bibinfo {author} {\bibfnamefont
  {P.}~\bibnamefont {Roushan}}, \bibinfo {author} {\bibfnamefont
  {A.}~\bibnamefont {Tranter}}, \bibinfo {author} {\bibfnamefont
  {N.}~\bibnamefont {Ding}},  \emph {et~al.},\ }\href@noop {} {\bibfield
  {journal} {\bibinfo  {journal} {Physical Review X}\ }\textbf {\bibinfo
  {volume} {6}},\ \bibinfo {pages} {031007} (\bibinfo {year}
  {2016})}\BibitemShut {NoStop}%
\bibitem [{\citenamefont {Peruzzo}\ \emph {et~al.}(2014)\citenamefont
  {Peruzzo}, \citenamefont {McClean}, \citenamefont {Shadbolt}, \citenamefont
  {Yung}, \citenamefont {Zhou}, \citenamefont {Love}, \citenamefont
  {Aspuru-Guzik},\ and\ \citenamefont {O’brien}}]{peruzzo2014variational}%
  \BibitemOpen
  \bibfield  {author} {\bibinfo {author} {\bibfnamefont {A.}~\bibnamefont
  {Peruzzo}}, \bibinfo {author} {\bibfnamefont {J.}~\bibnamefont {McClean}},
  \bibinfo {author} {\bibfnamefont {P.}~\bibnamefont {Shadbolt}}, \bibinfo
  {author} {\bibfnamefont {M.-H.}\ \bibnamefont {Yung}}, \bibinfo {author}
  {\bibfnamefont {X.-Q.}\ \bibnamefont {Zhou}}, \bibinfo {author}
  {\bibfnamefont {P.~J.}\ \bibnamefont {Love}}, \bibinfo {author}
  {\bibfnamefont {A.}~\bibnamefont {Aspuru-Guzik}}, \ and\ \bibinfo {author}
  {\bibfnamefont {J.~L.}\ \bibnamefont {O’brien}},\ }\href@noop {} {\bibfield
   {journal} {\bibinfo  {journal} {Nature Communications}\ }\textbf {\bibinfo
  {volume} {5}},\ \bibinfo {pages} {4213} (\bibinfo {year} {2014})}\BibitemShut
  {NoStop}%
\bibitem [{\citenamefont {McClean}\ \emph {et~al.}(2016)\citenamefont
  {McClean}, \citenamefont {Romero}, \citenamefont {Babbush},\ and\
  \citenamefont {Aspuru-Guzik}}]{mcclean2016theory}%
  \BibitemOpen
  \bibfield  {author} {\bibinfo {author} {\bibfnamefont {J.~R.}\ \bibnamefont
  {McClean}}, \bibinfo {author} {\bibfnamefont {J.}~\bibnamefont {Romero}},
  \bibinfo {author} {\bibfnamefont {R.}~\bibnamefont {Babbush}}, \ and\
  \bibinfo {author} {\bibfnamefont {A.}~\bibnamefont {Aspuru-Guzik}},\
  }\href@noop {} {\bibfield  {journal} {\bibinfo  {journal} {New Journal of
  Physics}\ }\textbf {\bibinfo {volume} {18}},\ \bibinfo {pages} {023023}
  (\bibinfo {year} {2016})}\BibitemShut {NoStop}%
\bibitem [{\citenamefont {Preskill}(2018)}]{preskill2018quantum}%
  \BibitemOpen
  \bibfield  {author} {\bibinfo {author} {\bibfnamefont {J.}~\bibnamefont
  {Preskill}},\ }\href@noop {} {\bibfield  {journal} {\bibinfo  {journal}
  {Quantum}\ }\textbf {\bibinfo {volume} {2}},\ \bibinfo {pages} {79} (\bibinfo
  {year} {2018})}\BibitemShut {NoStop}%
\bibitem [{\citenamefont {Shen}\ \emph {et~al.}(2017)\citenamefont {Shen},
  \citenamefont {Zhang}, \citenamefont {Zhang}, \citenamefont {Zhang},
  \citenamefont {Yung},\ and\ \citenamefont {Kim}}]{shen2017quantum}%
  \BibitemOpen
  \bibfield  {author} {\bibinfo {author} {\bibfnamefont {Y.}~\bibnamefont
  {Shen}}, \bibinfo {author} {\bibfnamefont {X.}~\bibnamefont {Zhang}},
  \bibinfo {author} {\bibfnamefont {S.}~\bibnamefont {Zhang}}, \bibinfo
  {author} {\bibfnamefont {J.-N.}\ \bibnamefont {Zhang}}, \bibinfo {author}
  {\bibfnamefont {M.-H.}\ \bibnamefont {Yung}}, \ and\ \bibinfo {author}
  {\bibfnamefont {K.}~\bibnamefont {Kim}},\ }\href@noop {} {\bibfield
  {journal} {\bibinfo  {journal} {Physical Review A}\ }\textbf {\bibinfo
  {volume} {95}},\ \bibinfo {pages} {020501} (\bibinfo {year}
  {2017})}\BibitemShut {NoStop}%
\bibitem [{\citenamefont {Kandala}\ \emph {et~al.}(2017)\citenamefont
  {Kandala}, \citenamefont {Mezzacapo}, \citenamefont {Temme}, \citenamefont
  {Takita}, \citenamefont {Brink}, \citenamefont {Chow},\ and\ \citenamefont
  {Gambetta}}]{kandala2017hardware}%
  \BibitemOpen
  \bibfield  {author} {\bibinfo {author} {\bibfnamefont {A.}~\bibnamefont
  {Kandala}}, \bibinfo {author} {\bibfnamefont {A.}~\bibnamefont {Mezzacapo}},
  \bibinfo {author} {\bibfnamefont {K.}~\bibnamefont {Temme}}, \bibinfo
  {author} {\bibfnamefont {M.}~\bibnamefont {Takita}}, \bibinfo {author}
  {\bibfnamefont {M.}~\bibnamefont {Brink}}, \bibinfo {author} {\bibfnamefont
  {J.~M.}\ \bibnamefont {Chow}}, \ and\ \bibinfo {author} {\bibfnamefont
  {J.~M.}\ \bibnamefont {Gambetta}},\ }\href@noop {} {\bibfield  {journal}
  {\bibinfo  {journal} {Nature}\ }\textbf {\bibinfo {volume} {549}},\ \bibinfo
  {pages} {242} (\bibinfo {year} {2017})}\BibitemShut {NoStop}%
\bibitem [{\citenamefont {Hempel}\ \emph {et~al.}(2018)\citenamefont {Hempel},
  \citenamefont {Maier}, \citenamefont {Romero}, \citenamefont {McClean},
  \citenamefont {Monz}, \citenamefont {Shen}, \citenamefont {Jurcevic},
  \citenamefont {Lanyon}, \citenamefont {Love}, \citenamefont {Babbush} \emph
  {et~al.}}]{hempel2018quantum}%
  \BibitemOpen
  \bibfield  {author} {\bibinfo {author} {\bibfnamefont {C.}~\bibnamefont
  {Hempel}}, \bibinfo {author} {\bibfnamefont {C.}~\bibnamefont {Maier}},
  \bibinfo {author} {\bibfnamefont {J.}~\bibnamefont {Romero}}, \bibinfo
  {author} {\bibfnamefont {J.}~\bibnamefont {McClean}}, \bibinfo {author}
  {\bibfnamefont {T.}~\bibnamefont {Monz}}, \bibinfo {author} {\bibfnamefont
  {H.}~\bibnamefont {Shen}}, \bibinfo {author} {\bibfnamefont {P.}~\bibnamefont
  {Jurcevic}}, \bibinfo {author} {\bibfnamefont {B.~P.}\ \bibnamefont
  {Lanyon}}, \bibinfo {author} {\bibfnamefont {P.}~\bibnamefont {Love}},
  \bibinfo {author} {\bibfnamefont {R.}~\bibnamefont {Babbush}},  \emph
  {et~al.},\ }\href@noop {} {\bibfield  {journal} {\bibinfo  {journal}
  {Physical Review X}\ }\textbf {\bibinfo {volume} {8}},\ \bibinfo {pages}
  {031022} (\bibinfo {year} {2018})}\BibitemShut {NoStop}%
\bibitem [{\citenamefont {Chiesa}\ \emph {et~al.}(2019)\citenamefont {Chiesa},
  \citenamefont {Tacchino}, \citenamefont {Grossi}, \citenamefont {Santini},
  \citenamefont {Tavernelli}, \citenamefont {Gerace},\ and\ \citenamefont
  {Carretta}}]{chiesa2019quantum}%
  \BibitemOpen
  \bibfield  {author} {\bibinfo {author} {\bibfnamefont {A.}~\bibnamefont
  {Chiesa}}, \bibinfo {author} {\bibfnamefont {F.}~\bibnamefont {Tacchino}},
  \bibinfo {author} {\bibfnamefont {M.}~\bibnamefont {Grossi}}, \bibinfo
  {author} {\bibfnamefont {P.}~\bibnamefont {Santini}}, \bibinfo {author}
  {\bibfnamefont {I.}~\bibnamefont {Tavernelli}}, \bibinfo {author}
  {\bibfnamefont {D.}~\bibnamefont {Gerace}}, \ and\ \bibinfo {author}
  {\bibfnamefont {S.}~\bibnamefont {Carretta}},\ }\href@noop {} {\bibfield
  {journal} {\bibinfo  {journal} {Nature Physics}\ }\textbf {\bibinfo {volume}
  {15}},\ \bibinfo {pages} {455} (\bibinfo {year} {2019})}\BibitemShut
  {NoStop}%
\bibitem [{\citenamefont {Francis}\ \emph {et~al.}(2020)\citenamefont
  {Francis}, \citenamefont {Freericks},\ and\ \citenamefont
  {Kemper}}]{francis2020quantum}%
  \BibitemOpen
  \bibfield  {author} {\bibinfo {author} {\bibfnamefont {A.}~\bibnamefont
  {Francis}}, \bibinfo {author} {\bibfnamefont {J.}~\bibnamefont {Freericks}},
  \ and\ \bibinfo {author} {\bibfnamefont {A.}~\bibnamefont {Kemper}},\
  }\href@noop {} {\bibfield  {journal} {\bibinfo  {journal} {Physical Review
  B}\ }\textbf {\bibinfo {volume} {101}},\ \bibinfo {pages} {014411} (\bibinfo
  {year} {2020})}\BibitemShut {NoStop}%
\bibitem [{\citenamefont {Norman}\ \emph {et~al.}(2005)\citenamefont {Norman},
  \citenamefont {Bishop}, \citenamefont {Jensen},\ and\ \citenamefont
  {Oddershede}}]{norman2005nonlinear}%
  \BibitemOpen
  \bibfield  {author} {\bibinfo {author} {\bibfnamefont {P.}~\bibnamefont
  {Norman}}, \bibinfo {author} {\bibfnamefont {D.~M.}\ \bibnamefont {Bishop}},
  \bibinfo {author} {\bibfnamefont {H.~J.~A.}\ \bibnamefont {Jensen}}, \ and\
  \bibinfo {author} {\bibfnamefont {J.}~\bibnamefont {Oddershede}},\
  }\href@noop {} {\bibfield  {journal} {\bibinfo  {journal} {The Journal of
  Chemical Physics}\ }\textbf {\bibinfo {volume} {123}},\ \bibinfo {pages}
  {194103} (\bibinfo {year} {2005})}\BibitemShut {NoStop}%
\bibitem [{\citenamefont {Mukamel}(2000)}]{mukamel2000multidimensional}%
  \BibitemOpen
  \bibfield  {author} {\bibinfo {author} {\bibfnamefont {S.}~\bibnamefont
  {Mukamel}},\ }\href@noop {} {\bibfield  {journal} {\bibinfo  {journal}
  {Annual Review of Physical Chemistry}\ }\textbf {\bibinfo {volume} {51}},\
  \bibinfo {pages} {691} (\bibinfo {year} {2000})}\BibitemShut {NoStop}%
\bibitem [{\citenamefont {Harrow}\ \emph {et~al.}(2009)\citenamefont {Harrow},
  \citenamefont {Hassidim},\ and\ \citenamefont {Lloyd}}]{harrow2009quantum}%
  \BibitemOpen
  \bibfield  {author} {\bibinfo {author} {\bibfnamefont {A.~W.}\ \bibnamefont
  {Harrow}}, \bibinfo {author} {\bibfnamefont {A.}~\bibnamefont {Hassidim}}, \
  and\ \bibinfo {author} {\bibfnamefont {S.}~\bibnamefont {Lloyd}},\
  }\href@noop {} {\bibfield  {journal} {\bibinfo  {journal} {Physical Review
  Letters}\ }\textbf {\bibinfo {volume} {103}},\ \bibinfo {pages} {150502}
  (\bibinfo {year} {2009})}\BibitemShut {NoStop}%
\bibitem [{\citenamefont {Ambainis}(2010)}]{ambainis2010variable}%
  \BibitemOpen
  \bibfield  {author} {\bibinfo {author} {\bibfnamefont {A.}~\bibnamefont
  {Ambainis}},\ }\href@noop {} {\bibfield  {journal} {\bibinfo  {journal}
  {arXiv:1010.4458}\ } (\bibinfo {year} {2010})}\BibitemShut {NoStop}%
\bibitem [{\citenamefont {Clader}\ \emph {et~al.}(2013)\citenamefont {Clader},
  \citenamefont {Jacobs},\ and\ \citenamefont
  {Sprouse}}]{clader2013preconditioned}%
  \BibitemOpen
  \bibfield  {author} {\bibinfo {author} {\bibfnamefont {B.~D.}\ \bibnamefont
  {Clader}}, \bibinfo {author} {\bibfnamefont {B.~C.}\ \bibnamefont {Jacobs}},
  \ and\ \bibinfo {author} {\bibfnamefont {C.~R.}\ \bibnamefont {Sprouse}},\
  }\href@noop {} {\bibfield  {journal} {\bibinfo  {journal} {Physical Review
  Letters}\ }\textbf {\bibinfo {volume} {110}},\ \bibinfo {pages} {250504}
  (\bibinfo {year} {2013})}\BibitemShut {NoStop}%
\bibitem [{\citenamefont {Childs}\ \emph {et~al.}(2017)\citenamefont {Childs},
  \citenamefont {Kothari},\ and\ \citenamefont {Somma}}]{childs2017quantum}%
  \BibitemOpen
  \bibfield  {author} {\bibinfo {author} {\bibfnamefont {A.~M.}\ \bibnamefont
  {Childs}}, \bibinfo {author} {\bibfnamefont {R.}~\bibnamefont {Kothari}}, \
  and\ \bibinfo {author} {\bibfnamefont {R.~D.}\ \bibnamefont {Somma}},\
  }\href@noop {} {\bibfield  {journal} {\bibinfo  {journal} {SIAM Journal on
  Computing}\ }\textbf {\bibinfo {volume} {46}},\ \bibinfo {pages} {1920}
  (\bibinfo {year} {2017})}\BibitemShut {NoStop}%
\bibitem [{\citenamefont {Suba{\c{s}}{\i}}\ \emph {et~al.}(2019)\citenamefont
  {Suba{\c{s}}{\i}}, \citenamefont {Somma},\ and\ \citenamefont
  {Orsucci}}]{subacsi2019quantum}%
  \BibitemOpen
  \bibfield  {author} {\bibinfo {author} {\bibfnamefont {Y.}~\bibnamefont
  {Suba{\c{s}}{\i}}}, \bibinfo {author} {\bibfnamefont {R.~D.}\ \bibnamefont
  {Somma}}, \ and\ \bibinfo {author} {\bibfnamefont {D.}~\bibnamefont
  {Orsucci}},\ }\href@noop {} {\bibfield  {journal} {\bibinfo  {journal}
  {Physical review letters}\ }\textbf {\bibinfo {volume} {122}},\ \bibinfo
  {pages} {060504} (\bibinfo {year} {2019})}\BibitemShut {NoStop}%
\bibitem [{\citenamefont {Xu}\ \emph {et~al.}(2019)\citenamefont {Xu},
  \citenamefont {Sun}, \citenamefont {Endo}, \citenamefont {Li}, \citenamefont
  {Benjamin},\ and\ \citenamefont {Yuan}}]{xu2019variational}%
  \BibitemOpen
  \bibfield  {author} {\bibinfo {author} {\bibfnamefont {X.}~\bibnamefont
  {Xu}}, \bibinfo {author} {\bibfnamefont {J.}~\bibnamefont {Sun}}, \bibinfo
  {author} {\bibfnamefont {S.}~\bibnamefont {Endo}}, \bibinfo {author}
  {\bibfnamefont {Y.}~\bibnamefont {Li}}, \bibinfo {author} {\bibfnamefont
  {S.~C.}\ \bibnamefont {Benjamin}}, \ and\ \bibinfo {author} {\bibfnamefont
  {X.}~\bibnamefont {Yuan}},\ }\href@noop {} {\bibfield  {journal} {\bibinfo
  {journal} {arXiv preprint arXiv:1909.03898}\ } (\bibinfo {year}
  {2019})}\BibitemShut {NoStop}%
\bibitem [{\citenamefont {Bravo-Prieto}\ \emph {et~al.}(2019)\citenamefont
  {Bravo-Prieto}, \citenamefont {LaRose}, \citenamefont {Cerezo}, \citenamefont
  {Subasi}, \citenamefont {Cincio},\ and\ \citenamefont
  {Coles}}]{bravo2019variational}%
  \BibitemOpen
  \bibfield  {author} {\bibinfo {author} {\bibfnamefont {C.}~\bibnamefont
  {Bravo-Prieto}}, \bibinfo {author} {\bibfnamefont {R.}~\bibnamefont
  {LaRose}}, \bibinfo {author} {\bibfnamefont {M.}~\bibnamefont {Cerezo}},
  \bibinfo {author} {\bibfnamefont {Y.}~\bibnamefont {Subasi}}, \bibinfo
  {author} {\bibfnamefont {L.}~\bibnamefont {Cincio}}, \ and\ \bibinfo {author}
  {\bibfnamefont {P.~J.}\ \bibnamefont {Coles}},\ }\href@noop {} {\bibfield
  {journal} {\bibinfo  {journal} {arXiv preprint arXiv:1909.05820}\ } (\bibinfo
  {year} {2019})}\BibitemShut {NoStop}%
\bibitem [{\citenamefont {Saad}(2003)}]{saad2003iterative}%
  \BibitemOpen
  \bibfield  {author} {\bibinfo {author} {\bibfnamefont {Y.}~\bibnamefont
  {Saad}},\ }\href@noop {} {\emph {\bibinfo {title} {Iterative methods for
  sparse linear systems}}},\ Vol.~\bibinfo {volume} {82}\ (\bibinfo
  {publisher} {Society for Industrial and Applied Mathematics},\ \bibinfo
  {year} {2003})\BibitemShut {NoStop}%
\bibitem [{\citenamefont {Aaronson}(2015)}]{aaronson2015read}%
  \BibitemOpen
  \bibfield  {author} {\bibinfo {author} {\bibfnamefont {S.}~\bibnamefont
  {Aaronson}},\ }\href@noop {} {\bibfield  {journal} {\bibinfo  {journal}
  {Nature Physics}\ }\textbf {\bibinfo {volume} {11}},\ \bibinfo {pages} {291}
  (\bibinfo {year} {2015})}\BibitemShut {NoStop}%
\bibitem [{\citenamefont {Childs}\ and\ \citenamefont
  {Wiebe}(2012)}]{Childs2012}%
  \BibitemOpen
  \bibfield  {author} {\bibinfo {author} {\bibfnamefont {A.~M.}\ \bibnamefont
  {Childs}}\ and\ \bibinfo {author} {\bibfnamefont {N.}~\bibnamefont {Wiebe}},\
  }\href {http://dl.acm.org/citation.cfm?id=2481569.2481570} {\bibfield
  {journal} {\bibinfo  {journal} {Quantum Info. Comput.}\ }\textbf {\bibinfo
  {volume} {12}},\ \bibinfo {pages} {901} (\bibinfo {year} {2012})}\BibitemShut
  {NoStop}%
\bibitem [{\citenamefont {Berry}\ \emph {et~al.}(2015)\citenamefont {Berry},
  \citenamefont {Childs}, \citenamefont {Cleve}, \citenamefont {Kothari},\ and\
  \citenamefont {Somma}}]{berry2015simulating}%
  \BibitemOpen
  \bibfield  {author} {\bibinfo {author} {\bibfnamefont {D.~W.}\ \bibnamefont
  {Berry}}, \bibinfo {author} {\bibfnamefont {A.~M.}\ \bibnamefont {Childs}},
  \bibinfo {author} {\bibfnamefont {R.}~\bibnamefont {Cleve}}, \bibinfo
  {author} {\bibfnamefont {R.}~\bibnamefont {Kothari}}, \ and\ \bibinfo
  {author} {\bibfnamefont {R.~D.}\ \bibnamefont {Somma}},\ }\href@noop {}
  {\bibfield  {journal} {\bibinfo  {journal} {Physical Review Letters}\
  }\textbf {\bibinfo {volume} {114}},\ \bibinfo {pages} {090502} (\bibinfo
  {year} {2015})}\BibitemShut {NoStop}%
\bibitem [{\citenamefont {Jordan}\ and\ \citenamefont
  {Wigner}(1928)}]{jordan1928pauli}%
  \BibitemOpen
  \bibfield  {author} {\bibinfo {author} {\bibfnamefont {P.}~\bibnamefont
  {Jordan}}\ and\ \bibinfo {author} {\bibfnamefont {E.}~\bibnamefont
  {Wigner}},\ }\href@noop {} {\bibfield  {journal} {\bibinfo  {journal} {Z.
  Phys}\ }\textbf {\bibinfo {volume} {47}},\ \bibinfo {pages} {14} (\bibinfo
  {year} {1928})}\BibitemShut {NoStop}%
\bibitem [{\citenamefont {Bravyi}\ and\ \citenamefont
  {Kitaev}(2002)}]{bravyi2002fermionic}%
  \BibitemOpen
  \bibfield  {author} {\bibinfo {author} {\bibfnamefont {S.~B.}\ \bibnamefont
  {Bravyi}}\ and\ \bibinfo {author} {\bibfnamefont {A.~Y.}\ \bibnamefont
  {Kitaev}},\ }\href@noop {} {\bibfield  {journal} {\bibinfo  {journal} {Annals
  of Physics}\ }\textbf {\bibinfo {volume} {298}},\ \bibinfo {pages} {210}
  (\bibinfo {year} {2002})}\BibitemShut {NoStop}%
\bibitem [{\citenamefont {Seeley}\ \emph {et~al.}(2012)\citenamefont {Seeley},
  \citenamefont {Richard},\ and\ \citenamefont {Love}}]{seeley2012bravyi}%
  \BibitemOpen
  \bibfield  {author} {\bibinfo {author} {\bibfnamefont {J.~T.}\ \bibnamefont
  {Seeley}}, \bibinfo {author} {\bibfnamefont {M.~J.}\ \bibnamefont {Richard}},
  \ and\ \bibinfo {author} {\bibfnamefont {P.~J.}\ \bibnamefont {Love}},\
  }\href@noop {} {\bibfield  {journal} {\bibinfo  {journal} {The Journal of
  Chemical Physics}\ }\textbf {\bibinfo {volume} {137}},\ \bibinfo {pages}
  {224109} (\bibinfo {year} {2012})}\BibitemShut {NoStop}%
\bibitem [{\citenamefont {Wecker}\ \emph {et~al.}(2015)\citenamefont {Wecker},
  \citenamefont {Hastings},\ and\ \citenamefont {Troyer}}]{wecker2015progress}%
  \BibitemOpen
  \bibfield  {author} {\bibinfo {author} {\bibfnamefont {D.}~\bibnamefont
  {Wecker}}, \bibinfo {author} {\bibfnamefont {M.~B.}\ \bibnamefont
  {Hastings}}, \ and\ \bibinfo {author} {\bibfnamefont {M.}~\bibnamefont
  {Troyer}},\ }\href@noop {} {\bibfield  {journal} {\bibinfo  {journal}
  {Physical Review A}\ }\textbf {\bibinfo {volume} {92}},\ \bibinfo {pages}
  {042303} (\bibinfo {year} {2015})}\BibitemShut {NoStop}%
\bibitem [{\citenamefont {Buhrman}\ \emph {et~al.}(2001)\citenamefont
  {Buhrman}, \citenamefont {Cleve}, \citenamefont {Watrous},\ and\
  \citenamefont {De~Wolf}}]{buhrman2001quantum}%
  \BibitemOpen
  \bibfield  {author} {\bibinfo {author} {\bibfnamefont {H.}~\bibnamefont
  {Buhrman}}, \bibinfo {author} {\bibfnamefont {R.}~\bibnamefont {Cleve}},
  \bibinfo {author} {\bibfnamefont {J.}~\bibnamefont {Watrous}}, \ and\
  \bibinfo {author} {\bibfnamefont {R.}~\bibnamefont {De~Wolf}},\ }\href@noop
  {} {\bibfield  {journal} {\bibinfo  {journal} {Physical Review Letters}\
  }\textbf {\bibinfo {volume} {87}},\ \bibinfo {pages} {167902} (\bibinfo
  {year} {2001})}\BibitemShut {NoStop}%
\bibitem [{\citenamefont {Garcia-Escartin}\ and\ \citenamefont
  {Chamorro-Posada}(2013)}]{garcia2013swap}%
  \BibitemOpen
  \bibfield  {author} {\bibinfo {author} {\bibfnamefont {J.~C.}\ \bibnamefont
  {Garcia-Escartin}}\ and\ \bibinfo {author} {\bibfnamefont {P.}~\bibnamefont
  {Chamorro-Posada}},\ }\href@noop {} {\bibfield  {journal} {\bibinfo
  {journal} {Physical Review A}\ }\textbf {\bibinfo {volume} {87}},\ \bibinfo
  {pages} {052330} (\bibinfo {year} {2013})}\BibitemShut {NoStop}%
\bibitem [{\citenamefont {Cai}\ \emph {et~al.}(2013)\citenamefont {Cai},
  \citenamefont {Weedbrook}, \citenamefont {Su}, \citenamefont {Chen},
  \citenamefont {Gu}, \citenamefont {Zhu}, \citenamefont {Li}, \citenamefont
  {Liu}, \citenamefont {Lu},\ and\ \citenamefont {Pan}}]{cai2013experimental}%
  \BibitemOpen
  \bibfield  {author} {\bibinfo {author} {\bibfnamefont {X.-D.}\ \bibnamefont
  {Cai}}, \bibinfo {author} {\bibfnamefont {C.}~\bibnamefont {Weedbrook}},
  \bibinfo {author} {\bibfnamefont {Z.-E.}\ \bibnamefont {Su}}, \bibinfo
  {author} {\bibfnamefont {M.-C.}\ \bibnamefont {Chen}}, \bibinfo {author}
  {\bibfnamefont {M.}~\bibnamefont {Gu}}, \bibinfo {author} {\bibfnamefont
  {M.-J.}\ \bibnamefont {Zhu}}, \bibinfo {author} {\bibfnamefont
  {L.}~\bibnamefont {Li}}, \bibinfo {author} {\bibfnamefont {N.-L.}\
  \bibnamefont {Liu}}, \bibinfo {author} {\bibfnamefont {C.-Y.}\ \bibnamefont
  {Lu}}, \ and\ \bibinfo {author} {\bibfnamefont {J.-W.}\ \bibnamefont {Pan}},\
  }\href@noop {} {\bibfield  {journal} {\bibinfo  {journal} {Physical Review
  Letters}\ }\textbf {\bibinfo {volume} {110}},\ \bibinfo {pages} {230501}
  (\bibinfo {year} {2013})}\BibitemShut {NoStop}%
\bibitem [{\citenamefont {Pan}\ \emph {et~al.}(2014)\citenamefont {Pan},
  \citenamefont {Cao}, \citenamefont {Yao}, \citenamefont {Li}, \citenamefont
  {Ju}, \citenamefont {Chen}, \citenamefont {Peng}, \citenamefont {Kais},\ and\
  \citenamefont {Du}}]{pan2014experimental}%
  \BibitemOpen
  \bibfield  {author} {\bibinfo {author} {\bibfnamefont {J.}~\bibnamefont
  {Pan}}, \bibinfo {author} {\bibfnamefont {Y.}~\bibnamefont {Cao}}, \bibinfo
  {author} {\bibfnamefont {X.}~\bibnamefont {Yao}}, \bibinfo {author}
  {\bibfnamefont {Z.}~\bibnamefont {Li}}, \bibinfo {author} {\bibfnamefont
  {C.}~\bibnamefont {Ju}}, \bibinfo {author} {\bibfnamefont {H.}~\bibnamefont
  {Chen}}, \bibinfo {author} {\bibfnamefont {X.}~\bibnamefont {Peng}}, \bibinfo
  {author} {\bibfnamefont {S.}~\bibnamefont {Kais}}, \ and\ \bibinfo {author}
  {\bibfnamefont {J.}~\bibnamefont {Du}},\ }\href@noop {} {\bibfield  {journal}
  {\bibinfo  {journal} {Physical Review A}\ }\textbf {\bibinfo {volume} {89}},\
  \bibinfo {pages} {022313} (\bibinfo {year} {2014})}\BibitemShut {NoStop}%
\bibitem [{\citenamefont {Barz}\ \emph {et~al.}(2014)\citenamefont {Barz},
  \citenamefont {Kassal}, \citenamefont {Ringbauer}, \citenamefont {Lipp},
  \citenamefont {Daki{\'c}}, \citenamefont {Aspuru-Guzik},\ and\ \citenamefont
  {Walther}}]{barz2014two}%
  \BibitemOpen
  \bibfield  {author} {\bibinfo {author} {\bibfnamefont {S.}~\bibnamefont
  {Barz}}, \bibinfo {author} {\bibfnamefont {I.}~\bibnamefont {Kassal}},
  \bibinfo {author} {\bibfnamefont {M.}~\bibnamefont {Ringbauer}}, \bibinfo
  {author} {\bibfnamefont {Y.~O.}\ \bibnamefont {Lipp}}, \bibinfo {author}
  {\bibfnamefont {B.}~\bibnamefont {Daki{\'c}}}, \bibinfo {author}
  {\bibfnamefont {A.}~\bibnamefont {Aspuru-Guzik}}, \ and\ \bibinfo {author}
  {\bibfnamefont {P.}~\bibnamefont {Walther}},\ }\href@noop {} {\bibfield
  {journal} {\bibinfo  {journal} {Scientific Reports}\ }\textbf {\bibinfo
  {volume} {4}},\ \bibinfo {pages} {6115} (\bibinfo {year} {2014})}\BibitemShut
  {NoStop}%
\bibitem [{\citenamefont {Zheng}\ \emph {et~al.}(2017)\citenamefont {Zheng},
  \citenamefont {Song}, \citenamefont {Chen}, \citenamefont {Xia},
  \citenamefont {Liu}, \citenamefont {Guo}, \citenamefont {Zhang},
  \citenamefont {Xu}, \citenamefont {Deng}, \citenamefont {Huang} \emph
  {et~al.}}]{zheng2017solving}%
  \BibitemOpen
  \bibfield  {author} {\bibinfo {author} {\bibfnamefont {Y.}~\bibnamefont
  {Zheng}}, \bibinfo {author} {\bibfnamefont {C.}~\bibnamefont {Song}},
  \bibinfo {author} {\bibfnamefont {M.-C.}\ \bibnamefont {Chen}}, \bibinfo
  {author} {\bibfnamefont {B.}~\bibnamefont {Xia}}, \bibinfo {author}
  {\bibfnamefont {W.}~\bibnamefont {Liu}}, \bibinfo {author} {\bibfnamefont
  {Q.}~\bibnamefont {Guo}}, \bibinfo {author} {\bibfnamefont {L.}~\bibnamefont
  {Zhang}}, \bibinfo {author} {\bibfnamefont {D.}~\bibnamefont {Xu}}, \bibinfo
  {author} {\bibfnamefont {H.}~\bibnamefont {Deng}}, \bibinfo {author}
  {\bibfnamefont {K.}~\bibnamefont {Huang}},  \emph {et~al.},\ }\href@noop {}
  {\bibfield  {journal} {\bibinfo  {journal} {Physical Review Letters}\
  }\textbf {\bibinfo {volume} {118}},\ \bibinfo {pages} {210504} (\bibinfo
  {year} {2017})}\BibitemShut {NoStop}%
\bibitem [{\citenamefont {Li}\ and\ \citenamefont
  {Benjamin}(2017)}]{li2017efficient}%
  \BibitemOpen
  \bibfield  {author} {\bibinfo {author} {\bibfnamefont {Y.}~\bibnamefont
  {Li}}\ and\ \bibinfo {author} {\bibfnamefont {S.~C.}\ \bibnamefont
  {Benjamin}},\ }\href@noop {} {\bibfield  {journal} {\bibinfo  {journal}
  {Physical Review X}\ }\textbf {\bibinfo {volume} {7}},\ \bibinfo {pages}
  {021050} (\bibinfo {year} {2017})}\BibitemShut {NoStop}%
\bibitem [{\citenamefont {Temme}\ \emph {et~al.}(2017)\citenamefont {Temme},
  \citenamefont {Bravyi},\ and\ \citenamefont {Gambetta}}]{temme2017error}%
  \BibitemOpen
  \bibfield  {author} {\bibinfo {author} {\bibfnamefont {K.}~\bibnamefont
  {Temme}}, \bibinfo {author} {\bibfnamefont {S.}~\bibnamefont {Bravyi}}, \
  and\ \bibinfo {author} {\bibfnamefont {J.~M.}\ \bibnamefont {Gambetta}},\
  }\href@noop {} {\bibfield  {journal} {\bibinfo  {journal} {Physical review
  letters}\ }\textbf {\bibinfo {volume} {119}},\ \bibinfo {pages} {180509}
  (\bibinfo {year} {2017})}\BibitemShut {NoStop}%
\bibitem [{\citenamefont {McArdle}\ \emph {et~al.}(2019)\citenamefont
  {McArdle}, \citenamefont {Yuan},\ and\ \citenamefont
  {Benjamin}}]{mcardle2019error}%
  \BibitemOpen
  \bibfield  {author} {\bibinfo {author} {\bibfnamefont {S.}~\bibnamefont
  {McArdle}}, \bibinfo {author} {\bibfnamefont {X.}~\bibnamefont {Yuan}}, \
  and\ \bibinfo {author} {\bibfnamefont {S.}~\bibnamefont {Benjamin}},\
  }\href@noop {} {\bibfield  {journal} {\bibinfo  {journal} {Physical review
  letters}\ }\textbf {\bibinfo {volume} {122}},\ \bibinfo {pages} {180501}
  (\bibinfo {year} {2019})}\BibitemShut {NoStop}%
\bibitem [{\citenamefont {Kandala}\ \emph {et~al.}(2019)\citenamefont
  {Kandala}, \citenamefont {Temme}, \citenamefont {C{\'o}rcoles}, \citenamefont
  {Mezzacapo}, \citenamefont {Chow},\ and\ \citenamefont
  {Gambetta}}]{kandala2019error}%
  \BibitemOpen
  \bibfield  {author} {\bibinfo {author} {\bibfnamefont {A.}~\bibnamefont
  {Kandala}}, \bibinfo {author} {\bibfnamefont {K.}~\bibnamefont {Temme}},
  \bibinfo {author} {\bibfnamefont {A.~D.}\ \bibnamefont {C{\'o}rcoles}},
  \bibinfo {author} {\bibfnamefont {A.}~\bibnamefont {Mezzacapo}}, \bibinfo
  {author} {\bibfnamefont {J.~M.}\ \bibnamefont {Chow}}, \ and\ \bibinfo
  {author} {\bibfnamefont {J.~M.}\ \bibnamefont {Gambetta}},\ }\href@noop {}
  {\bibfield  {journal} {\bibinfo  {journal} {Nature}\ }\textbf {\bibinfo
  {volume} {567}},\ \bibinfo {pages} {491} (\bibinfo {year}
  {2019})}\BibitemShut {NoStop}%
\bibitem [{\citenamefont {Torlai}\ \emph {et~al.}(2020)\citenamefont {Torlai},
  \citenamefont {Mazzola}, \citenamefont {Carleo},\ and\ \citenamefont
  {Mezzacapo}}]{torlai2020precise}%
  \BibitemOpen
  \bibfield  {author} {\bibinfo {author} {\bibfnamefont {G.}~\bibnamefont
  {Torlai}}, \bibinfo {author} {\bibfnamefont {G.}~\bibnamefont {Mazzola}},
  \bibinfo {author} {\bibfnamefont {G.}~\bibnamefont {Carleo}}, \ and\ \bibinfo
  {author} {\bibfnamefont {A.}~\bibnamefont {Mezzacapo}},\ }\href@noop {}
  {\bibfield  {journal} {\bibinfo  {journal} {Physical Review Research}\
  }\textbf {\bibinfo {volume} {2}},\ \bibinfo {pages} {022060} (\bibinfo {year}
  {2020})}\BibitemShut {NoStop}%
\bibitem [{\citenamefont {Ament}\ \emph {et~al.}(2011)\citenamefont {Ament},
  \citenamefont {Van~Veenendaal}, \citenamefont {Devereaux}, \citenamefont
  {Hill},\ and\ \citenamefont {Van Den~Brink}}]{ament2011resonant}%
  \BibitemOpen
  \bibfield  {author} {\bibinfo {author} {\bibfnamefont {L.~J.}\ \bibnamefont
  {Ament}}, \bibinfo {author} {\bibfnamefont {M.}~\bibnamefont
  {Van~Veenendaal}}, \bibinfo {author} {\bibfnamefont {T.~P.}\ \bibnamefont
  {Devereaux}}, \bibinfo {author} {\bibfnamefont {J.~P.}\ \bibnamefont {Hill}},
  \ and\ \bibinfo {author} {\bibfnamefont {J.}~\bibnamefont {Van Den~Brink}},\
  }\href@noop {} {\bibfield  {journal} {\bibinfo  {journal} {Reviews of Modern
  Physics}\ }\textbf {\bibinfo {volume} {83}},\ \bibinfo {pages} {705}
  (\bibinfo {year} {2011})}\BibitemShut {NoStop}%
\bibitem [{\citenamefont {Kramers}\ and\ \citenamefont
  {Heisenberg}(1925)}]{kramers1925streuung}%
  \BibitemOpen
  \bibfield  {author} {\bibinfo {author} {\bibfnamefont {H.~A.}\ \bibnamefont
  {Kramers}}\ and\ \bibinfo {author} {\bibfnamefont {W.}~\bibnamefont
  {Heisenberg}},\ }\href@noop {} {\bibfield  {journal} {\bibinfo  {journal} {Z.
  Phys.}\ }\textbf {\bibinfo {volume} {31}},\ \bibinfo {pages} {681} (\bibinfo
  {year} {1925})}\BibitemShut {NoStop}%
\bibitem [{\citenamefont {Kosugi}\ and\ \citenamefont
  {Matsushita}(2020)}]{PhysRevResearch.2.033043}%
  \BibitemOpen
  \bibfield  {author} {\bibinfo {author} {\bibfnamefont {T.}~\bibnamefont
  {Kosugi}}\ and\ \bibinfo {author} {\bibfnamefont {Y.-i.}\ \bibnamefont
  {Matsushita}},\ }\href {\doibase 10.1103/PhysRevResearch.2.033043} {\bibfield
   {journal} {\bibinfo  {journal} {Phys. Rev. Research}\ }\textbf {\bibinfo
  {volume} {2}},\ \bibinfo {pages} {033043} (\bibinfo {year}
  {2020})}\BibitemShut {NoStop}%
\end{thebibliography}%

\end{document}